\documentstyle[12pt]{article}
\def\C{{\rm\kern.24em \vrule width.02em 
 height1.4ex depth-.05ex
\kern-.26em C}}
\def\R{{\rm I\kern-.20em R}}
\def\Z{{\rm\kern.26em \vrule width.02em height0.5ex depth0ex
 \kern.04em
 \vrule  width.02em height1.47ex depth-1ex \kern-.34em Z}}
\def\N{{\rm I\kern-.20em N}}
\def\Q{{\rm\kern.24em \vrule width.02em 
 height1.4ex depth-.05ex \kern-.26em Q}}
\begin{document}
\begin{titlepage}
 .
\vskip 3.5cm
\begin{center}
  {\bf \Large Symbolic Computation of Conservation Laws \\
              of Nonlinear Partial Differential Equations \\
              in Multi-dimensions} \\
\vskip 2.0cm
                          Willy Hereman \\
          Department of Mathematical and Computer Sciences \\
                      Colorado School of Mines \\
                   Golden, CO 80401-1887, U.S.A. \\
\end{center}
\vskip 2cm
\centerline{Dedicated to Ryan Sayers (1982-2003)}
\vskip 3cm
\begin{center}
{\rm Research Supported in Part by the National Science Foundation (NSF) 
under Grants Nos.\ DMS-9732069, DMS-9912293, and CCR-9901929.} \\
\vskip 1cm
\end{center}
\begin{center}
To appear in: Thematic Issue on ``Mathematical Methods and Symbolic 
Calculation in Chemistry and Chemical Biology'' of the 
International Journal of Quantum Chemistry. 
Eds.: Michael Barnett and Frank Harris (2006). 
\end{center}
%
%
%
%
%
%
\end{titlepage}
\vfill
\newpage
\section*{Abstract}
%
%
A direct method for the computation of polynomial conservation laws of 
polynomial systems of nonlinear partial differential equations (PDEs) in 
multi-dimensions is presented.
The method avoids advanced differential-geometric tools. 
Instead, it is solely based on calculus, variational calculus, and linear 
algebra.

Densities are constructed as linear combinations of scaling homogeneous 
terms with undetermined coefficients.
The variational derivative (Euler operator) is used to compute the 
undetermined coefficients. 
The homotopy operator is used to compute the fluxes.

The method is illustrated with nonlinear PDEs describing wave phenomena 
in fluid dynamics, plasma physics, and quantum physics.
For PDEs with parameters, the method determines the conditions on the 
parameters so that a sequence of conserved densities might exist.
The existence of a large number of conservation laws is a predictor for 
complete integrability.
The method is algorithmic, applicable to a variety of PDEs, and can be 
implemented in computer algebra systems such as 
{\it Mathematica}, {\it Maple}, and {\it REDUCE}. 
%
%
\section{Introduction}
\label{introduction}
%
%
Nonlinear partial differential equations (PDEs) that admit conservation laws 
arise in many disciplines of the applied sciences including 
physical chemistry, fluid mechanics, particle and quantum physics, 
plasma physics, elasticity, gas dynamics, electromagnetism, 
magneto-hydro-dynamics, nonlinear optics, and the bio-sciences. 
Conservation laws are fundamental laws of physics.
They maintain that a certain quantity, e.g.\ momentum, mass (matter), electric 
charge, or energy, will not change with time during physical processes.
Often the PDE itself is a conservation law, e.g.\ the continuity equation 
relating charge to current.

As shown in \cite{MBetal04} and the articles in this issue, computer algebra 
systems (CAS) like {\it Mathematica}, {\it Maple}, and {\it REDUCE}, 
are useful to tackle computational problems in chemistry.
Finding closed-form conservation laws of nonlinear PDEs is a nice example.
Using CAS interactively, we could make a judicious guess (ansatz) and find a 
few simple densities and fluxes.
Yet, that approach is fruitless for complicated systems with nontrivial 
conservation laws.
Furthermore, completely integrable PDEs \cite{MAandPCbook91,MAandHSbook81} 
admit infinitely many independent conservation laws.
Computing them is a challenging task. 
It involves tedious computations which are prone to error if done with pen 
and paper.
The most famous example is the Korteweg-de Vries (KdV) equation from 
soliton theory \cite{MAandPCbook91,MAandHSbook81} which describes water 
waves in shallow water, ion-acoustic waves in plasmas, etc.\
Our earlier work \cite{PAthesis03,UGandWHjsc97} dealt primarily with the 
symbolic computation of conservation laws of completely integrable PDEs 
in $(1+1)$ dimensions (with independent variables $x$ and $t).$
In this paper we present a symbolic method that covers PDEs in 
multi-dimensions (e.g.\ $x,y,z,$ and $t),$ irrespective of their complete 
integrability.
As before, our approach relies on the concept of dilation (scaling) 
invariance which limits it to polynomial conserved densities of polynomial 
PDEs in evolution form.

There are many reasons to compute conserved densities and fluxes of PDEs 
explicitly.  
Invariants often lead to new discoveries as was the case in 
soliton theory.
We may want to verify if conserved quantities of physical importance 
(e.g.\ momentum, energy, Hamiltonians, entropy, density, charge) are 
intact after constitutive relations have been added to close a system.
For PDEs with arbitrary parameters we may wish to compute conditions on the 
parameters so that the model admits conserved quantities. 
Conserved densities also facilitate the study of qualitative properties of 
PDEs, such as bi- or tri-Hamiltonian structures.
They often guide the choice of solution methods or reveal the nature of 
special solutions.
For example, an infinite sequence of conserved densities assures complete 
integrability \cite{MAandPCbook91} of the PDE, i.e.\ solvability by the 
Inverse Scattering Transform \cite{MAandPCbook91} and the existence of 
solitons \cite{MAandHSbook81}.

Conserved densities aid in the design of numerical solvers for PDEs 
\cite{JMS1982}.
Indeed, semi-discretizations that conserve discrete conserved quantities 
lead to stable numerical schemes (i.e.\ free of nonlinear instabilities 
and blowup). 
While solving differential-difference equations (DDEs), which arise in 
nonlinear lattices and as semi-discretizations of PDEs, one should check 
that their conserved quantities indeed remain unchanged. 
Capitalizing on the analogy between PDEs and DDEs, the techniques presented 
in this paper have been adapted to DDEs 
\cite{HEthesis03,UGandWHpd98,WHetalbook05,WHetalcrm04,MHandWHprsa03} and 
fully discretized lattices \cite{MGetalcpc02,MGetal04}.

There are various methods (see \cite{WHetalbook05}) to compute conservation 
laws of nonlinear PDEs.
A common approach relies on the link between conservation laws and symmetries 
as stated in Noether's theorem \cite{IAbook04,IKandAVbook98,PObook93}.
However, the computation of generalized (variational) symmetries, 
though algorithmic, is as daunting a task as the direct computation of 
conservation laws. 
Nonetheless, we draw the reader's attention to \verb|DE_APPLS|, a package 
for constructing conservation laws from symmetries available within 
{\it Vessiot} \cite{IAvessiothandbook04}, a general purpose suite of 
{\it Maple} packages for computations on jet spaces.
Other methods circumvent the existence of a variational principle 
\cite{SA03,SAandGB02,TW02,TWetal03} 
but still rely on the symbolic solution of a determining system of PDEs.  
Despite their power, only a few of the above methods have been fully 
implemented in CAS (see \cite{UGandWHjsc97,WHetalbook05,TW02}).

We purposely avoid Noether's theorem, pre-knowledge of symmetries, and a 
Lagrangian formulation. 
Neither do we use differential forms or advanced differential-geometric tools.
Instead, we present and implement our tools in the language of calculus, 
linear algebra, and variational calculus. 
Our down-to-earth calculus formulas are transparent, easy to use by 
scientists and engineers, and are readily adaptable to nonlinear DDEs 
(not covered in {\it Vessiot}).

To design a reliable algorithm, we must compute both densities and fluxes.
For the latter, we need to invert the divergence operator which requires 
the integration (by parts) of an expression involving arbitrary functions. 
That is where the {\it homotopy operator} comes into play. 
Indeed, the homotopy operator reduces that problem to a standard 
one-dimensional integration with respect to a single auxiliary parameter.
One of the first\footnote{We refer the reader to \cite[p.\ 374]{PObook93} 
for a brief history of the homotopy operator.} uses of the homotopy operator 
in the context of conservation laws can be found in \cite{SAandGB02}.
The homotopy operator is a universal, yet little known, tool that can be 
applied to many problems in which integration by parts in multi-variables 
is needed.
A literature search revealed that homotopy operators are used in integrability
testing and inversion problems involving PDEs, DDEs, lattices, and 
beyond \cite{WHetalbook05}.
Assuming the reader is unfamiliar with homotopy operators, 
we ``demystify'' the homotopy formulas and make them ready for use in 
nonlinear sciences.

The particular application in this paper is computation of conservation laws
for which our algorithm proceeds as follows: 
build a candidate density as a linear combination (with undetermined 
coefficients) of ``building blocks'' that are homogeneous under the 
scaling symmetry of the PDE. 
If no such symmetry exists, construct one by introducing parameters with 
scaling.
Subsequently, use the conservation equation and the variational derivative 
to derive a linear algebraic system for the undetermined coefficients.
After the system is solved, use the homotopy operator to compute the flux. 
Implementations for $(1+1)$-dimensional PDEs \cite{UGandWHjsc97,UGandWHpd98} 
in {\it Mathematica} and {\it Maple} can be downloaded from 
\cite{BDwebsite04,WHwebsite04}.
{\it Mathematica} code that automates the computations for PDEs in 
multi-dimensions is being designed.

This paper is organized as follows. 
We establish some connections with vector calculus in 
Section~\ref{vectorcalculus}.
In Section~\ref{examples} we list the nonlinear PDEs that will be used 
throughout the paper:
the Korteweg-de Vries and Boussinesq equations from soliton theory
\cite{MAandPCbook91,MAandHSbook81}, 
the Landau-Lifshitz equation for Heisenberg's ferromagnet \cite{LFandLT87} 
and a system of shallow water wave equations for ocean waves \cite{PDpf03}. 
Sections~\ref{dilationinvariance} and~\ref{conservationlaws} cover the 
dilation invariance and conservation laws of those four examples.
In Section~\ref{tools}, we introduce and apply the tools from the calculus 
of variations. 
Formulas for the variational derivative in multi-dimensions are given 
in Section~\ref{zeroeuler}, where we apply the Euler operator for testing 
the exactness of various expressions.
Removal of divergence-equivalent terms with the Euler operator is 
discussed in Section~\ref{algorithm}.
The higher Euler operators and homotopy operators in 1D and 2D are in 
Sections~\ref{highereuler} and~\ref{homotopy}. 
In the latter section, we apply the homotopy operator to integrate by parts 
and to invert of the divergence operator.
In Section~\ref{applications} we present our three-step algorithm to 
compute conservation laws and apply it to the 
KdV equation (Section~\ref{applkdv}), the Boussinesq equation 
(Section~\ref{applboussinesq}), the Landau-Lifshitz equation 
(Section~\ref{applll}), and the shallow water wave equations 
(Section~\ref{applsww}).
Conclusions are drawn in Section~\ref{conclusions}.
\section{Connections with Vector Calculus}
\label{vectorcalculus}
%
%
We address a few issues in multivariate calculus which complement the 
material in \cite{JCetal02}:
(i) To determine whether or not a vector field ${\bf F}$ is 
{\it conservative}, i.e.\ ${\mathbf F} = {\mbox {\boldmath $\nabla$}} f$ 
for some scalar field $f,$ one must verify that ${\mathbf F}$ is 
{\it irrotational} or {\it curl free}, that is 
${\mbox {\boldmath $\nabla$}} \times {\mathbf F} = {\mathbf 0}.$
The cross $(\times)$ denotes the Euclidean cross product.
The field $f$ can be computed via standard integrations 
\cite[p.~518,~522]{JMandATbook88}.
\vskip 1pt
\noindent
(ii) To test if ${\mathbf F}$ is the curl of some vector field ${\mathbf G},$ 
one must check that ${\mathbf F}$ is {\it incompressible} or 
{\it divergence free}, i.e.\ 
${\mbox {\boldmath $\nabla$}} \cdot {\mathbf F} = 0.$ 
The dot $(\cdot)$ denotes the Euclidean inner product. 
The components of ${\mathbf G}$ result from solving a coupled system of 
first-order PDEs \cite[p.\ 526]{JMandATbook88}.
\vskip 2pt
\noindent
(iii) To verify whether or not a scalar field $f$ is the divergence of 
some vector function ${\mathbf F},$ no theorem from vector calculus
comes to the rescue.
Furthermore, the computation of ${\mathbf F}$ such that 
$f = {\mbox {\boldmath $\nabla$}} \cdot {\mathbf F}$ is a nontrivial matter
which requires the use of the homotopy operator or Hodge decomposition 
\cite{JCetal02}.
In single variable calculus, it amounts to computing the primitive 
$F = \int f \, dx.$

In multivariate calculus, all scalar fields $f,$ including the components 
$F_i$ of vector fields ${\mathbf F} = (F_1,F_2,F_3),$ are functions of the 
independent variables $(x,y,z).$ 
In differential geometry the above issues are addressed in much greater 
generality. 
The functions $f$ and $F_i$ can now depend on arbitrary functions 
$u(x,y,z),v(x,y,z),$ etc.\ and their mixed derivatives 
(up to a fixed order) with respect to $(x,y,z).$ 
Such functions are called {\it differential functions} \cite{PObook93}.
As one might expect, carrying out the gradient-, curl-, or divergence-test 
requires advanced algebraic machinery. 
For example, to test whether or not 
$f = {\mbox {\boldmath $\nabla$}} \cdot {\mathbf F}$ requires the use of the  
variational derivative (Euler operator) in 2D or 3D as we will show
in Section~\ref{zeroeuler}.
The actual computation of ${\mathbf F}$ requires integration by parts. 
That is where the higher Euler and homotopy operators enter the picture 
(see Sections~\ref{highereuler} and~\ref{homotopy}).

At the moment, no major CAS have reliable routines for integrating 
expressions involving arbitrary functions and their derivatives.
As far as we know, CAS offer no functions to test if a differential function 
is a divergence.
Routines to symbolically invert the total divergence are certainly lacking.
In Section~\ref{tools} we will present these tools and apply them in 
Section~\ref{applications}.
%
%
\section{Examples of Nonlinear PDEs}
\label{examples}
{\bf Definition}:
We consider a nonlinear system of {\it evolution} equations in $(3+1)$ 
dimensions, 
\begin{equation}
\label{continuoussystem}
{\bf u}_t = {\bf G}({\bf u}, {\bf u}_x, {\bf u}_y, {\bf u}_z,
{\bf u}_{2x}, {\bf u}_{2y}, {\bf u}_{2z}, 
{\bf u}_{xy}, {\bf u}_{xz}, {\bf u}_{yz}, \dots),
\end{equation}
where ${\bf x} \!=\! (x,y,z)$ and $t$ are space and time variables.
The vector ${\bf u}(x,y,z,t)$ has $N$ components $u_i.$ 
In the examples we denote the components of ${\bf u}$ by $u,v,w,$ etc..
Throughout the paper we use the subscript notation for partial derivatives,
\begin{equation} 
\label{notations}
{\bf u}_t = \frac{\partial {\bf u}}{\partial t}, \quad
{\bf u}_{2x} = {\bf u}_{xx} = \frac{\partial^2 {\bf u}}{\partial x^2}, \quad
{\bf u}_{2x3y} = {\bf u}_{xxyyy} 
= \frac{\partial^5 {\bf u}}{\partial x^2 \partial y^3}, \quad
{\bf u}_{xy} = \frac{\partial^2 {\bf u}}{\partial x\partial y}, \quad etc..
\end{equation}
We assume that ${\bf G}$ is smooth \cite{JCetal02} and does not explicitly 
depend on ${\bf x}$ and $t.$
There are no restrictions on the number of components, order, and degree of 
nonlinearity of ${\bf G}.$ 

We will predominantly work with polynomial systems, although systems 
involving one transcendental nonlinearity can also be handled 
\cite{PAthesis03,WHetalbook05}.
If parameters are present in (\ref{continuoussystem}), they will be denoted 
by lower-case Greek letters. 
Throughout this paper we work with the following four prototypical PDEs: 
\vskip 3pt
\noindent
{\bf Example 1}:
The ubiquitous {\rm Korteweg-de Vries (KdV) equation} 
\cite{MAandPCbook91,MAandHSbook81}
\begin{equation}
\label{kdv}
u_t + u u_x + u_{3x} = 0, 
\end{equation}
for $u(x,t)$ describes unidirectional shallow water waves and ion-acoustic 
waves in plasmas.
\vskip 4pt
\noindent
{\bf Example 2}:
The wave equation, 
\begin{equation} 
\label{boussinesq}
u_{tt} - u_{2x} + 3 u u_{2x} + 3 u_x^2 + \alpha u_{4x} = 0,
\end{equation}
for $u(x,t)$ with real parameter $\alpha,$ was proposed by Boussinesq to 
describe surface waves in shallow water \cite{MAandPCbook91}.
For what follows, we rewrite (\ref{boussinesq}) as a system of evolution 
equations,
\begin{equation}
\label{boussinesqsys}
u_t + v_x = 0, \quad
v_t + u_x - 3 u u_x - \alpha u_{3x} = 0,
\end{equation}
where $v(x,t)$ is an auxiliary dependent variable.
\vskip 4pt
\noindent
\vskip 2pt
\noindent
{\bf Example 3}:
The classical Landau-Lifshitz (LL) equation, 
\begin{equation}
\label{landaulifshitz}
{\bf S}_t = {\bf S} \times \Delta {\bf S} + {\bf S} \times D {\bf S},
\end{equation}
without external magnetic field, models nonlinear spin waves in a 
continuous Heisenberg ferromagnet \cite{LFandLT87,MSandHF93}.
The spin vector ${\bf S}(x,t)$ has real components 
$(u(x,t),v(x,t),w(x,t));$
$\Delta \!=\! {\mbox {\boldmath $\nabla$}^2}$ is the Laplacian, and 
$D \!=\! {\rm diag}(\alpha,\beta,\gamma)$ is a diagonal matrix with real 
coupling constants $\alpha,\beta,$ and $\gamma$ along the downward diagonal.
Splitting (\ref{landaulifshitz}) into components we get
\begin{eqnarray}
\label{heisenberg}
u_t &=& v w_{2x} - w v_{2x} + (\gamma - \beta) v w, 
\nonumber \\
v_t &=& w u_{2x} - u w_{2x} + (\alpha - \gamma) u w, 
\\
w_t &=& u v_{2x} - v u_{2x} + (\beta - \alpha) u v.
\nonumber
\end{eqnarray}
The second and third equations can be obtained from the first by 
{\it cyclic} permutations
\begin{equation}
\label{cyclicheisenberg}
(u \rightarrow v, v \rightarrow w, w \rightarrow u) 
\quad {\rm and} \quad
(\alpha \rightarrow \beta,\beta \rightarrow \gamma,\gamma \rightarrow \alpha).
\end{equation}
We take advantage of the cyclic nature of (\ref{heisenberg}) in the 
computations in Section~\ref{applll}.
\vskip 2pt
\noindent 
{\bf Example 4}:
The (2+1)-dimensional {\rm shallow-water wave (SWW) equations},
\begin{eqnarray}
\label{swwvector}
&& {\bf u}_t + ({\bf u} {\bf \cdot} {\mbox {\boldmath $\nabla$}} ) {\bf u} 
   + 2 {\mbox{\boldmath $\Omega$}} \times {\bf u} 
   = - {\mbox {\boldmath $\nabla$}} (h \theta ) 
     + {\textstyle \frac{1}{2}} h {\mbox {\boldmath $\nabla$}} \theta ,
\nonumber \\
&& 
h_t + {\mbox {\boldmath $\nabla$}} {\bf \cdot} (h {\bf u}) = 0, 
\quad
\theta_t + {\bf u} {\bf \cdot} ({\mbox {\boldmath $\nabla$}} \theta) = 0, 
\end{eqnarray}
describe waves in the ocean using layered models \cite{PDpf03}.
Vectors ${\bf u} = u(x,y,t) {\bf i} + v(x,y,t) {\bf j}$ and 
${\mbox {\boldmath $\Omega$}} = \Omega {\bf k}$ 
are the fluid and angular velocities.
${\mbox {\boldmath $\nabla$}} \!=\! 
\frac{\partial}{\partial x} {\bf i} + \frac{\partial}{\partial y} {\bf j}$
is the gradient operator and 
${\bf i}$, ${\bf j},$ and ${\bf k}$ are unit vectors along the $x$, $y,$ and 
$z$-axes. 
$\theta(x,y,t)$ is the horizontally varying potential temperature field
and $h(x,y,t)$ is the layer depth.
System (\ref{swwvector}) can be written as
\begin{eqnarray}
\label{sww}
&&\!\!\!\!\!\!\!\!\!\!\! u_t + u u_x + v u_y - 2 \Omega v + 
{\textstyle \frac{1}{2}} h \theta_x + \theta h_x \!=\! 0, 
\;
v_t + u v_x + v v_y + 2 \Omega u + 
{\textstyle \frac{1}{2}} h \theta_y + \theta h_y \!=\! 0, 
\nonumber \\
&&\!\!\!\!\!\!\!\!\!\!\! 
h_t + h u_x + u h_x + h v_y + v h_y \!=\! 0,
\quad
\theta_t + u \theta_x + v \theta_y \!=\! 0.
\end{eqnarray}
\section{Key Concept: Dilation or Scaling Invariance}
\label{dilationinvariance}
{\bf Definition}:
System (\ref{continuoussystem}) is {\it dilation invariant} if it does 
not change under a scaling symmetry.
\vskip 3pt
\noindent
{\bf Example}: 
The KdV equation (\ref{kdv}) is dilation invariant under the scaling symmetry
\begin{equation}
\label{kdvscale}
(x, t, u) \rightarrow (\lambda^{-1} x, \lambda^{-3} t, \lambda^2 u), 
\end{equation}
where $\lambda$ is an arbitrary parameter.
Indeed, apply the chain rule to verify that $\lambda^5$ factors out.
\vskip 3pt
\noindent
{\bf Definition}:
The {\it weight}\footnote{The weights are the remnants of physical units 
after non-dimensionalization of the PDE.}, $W,$
of a variable is the exponent in $\lambda^p$ which multiplies the variable.
\vskip 3pt
\noindent
{\bf Example}:
We will always replace $x$ by $\lambda^{-1} x.$  
Thus, $W(x) = -1$ or $W({\partial/\partial x}) = 1.$ 
In view of (\ref{kdvscale}), $W({\partial/\partial t}) = 3$ and $W(u) = 2$ 
for the KdV equation (\ref{kdv}).
\vskip 2pt
\noindent
{\bf Definition}:
The {\it rank} of a monomial is defined as the total weight of the monomial. 
An expression is {\it uniform in rank} if its monomial terms have equal rank.
\vskip 3pt
\noindent
{\bf Example}:
All monomials in (\ref{kdv}) have rank $5.$
Thus, (\ref{kdv}) is uniform in rank with respect to (\ref{kdvscale}).
\vskip 1pt
\noindent
Weights of dependent variables and weights of 
${\partial/\partial x}, {\partial/\partial y},$ etc.\ 
are assumed to be non-negative and rational.
Ranks must be positive integer or rational numbers.
The ranks of the equations in (\ref{continuoussystem}) may differ from 
each other. 

Conversely, requiring {\rm uniformity in rank} for each equation in 
(\ref{continuoussystem}) we can compute the weights, and thus the scaling 
symmetry, with linear algebra. 
\vskip 3pt
\noindent
{\bf Example}:
Indeed, for the KdV equation (\ref{kdv}) we have 
\begin{equation}
\label{kdvweightequations}
W(u) + W({\partial/\partial t}) = 2 W(u) + 1 = W(u) + 3,
\end{equation}
where we used $W({\partial/\partial x}) = 1.$ 
Solving (\ref{kdvweightequations}) yields 
\begin{equation}
\label{kdvweights}
W(u) = 2, \quad W({\partial/\partial t}) = 3.
\end{equation} 
This means that $x$ scales with $\lambda^{-1}, t$ with $\lambda^{-3},$
and $u$ with $\lambda^2,$ as claimed in (\ref{kdvscale}).
\vskip 4pt
\noindent
Dilation symmetries, which are special Lie-point symmetries \cite{PObook93}, 
are common to many nonlinear PDEs. 
However, non-uniform PDEs can be made uniform by giving appropriate weights
to parameters that appear in the PDEs, or, if necessary, by extending 
the set of dependent variables with {\it auxiliary} parameters with 
appropriate weights. 
Upon completion of the computations we can set the auxiliary parameters 
equal to $1.$
\vskip 4pt
\noindent
{\bf Example}:
The Boussinesq system (\ref{boussinesqsys}) is not uniform in rank 
because the terms $u_x$ and $\alpha u_{3x}$ lead to a contradiction in 
the weight equations.
To circumvent the problem we introduce an auxiliary parameter $\beta$ 
with (unknown) weight, and replace (\ref{boussinesqsys}) by 
\begin{eqnarray} 
\label{boussinesqsystem}
u_t + v_x &=& 0,
\nonumber \\
v_t + \beta u_x - 3 u u_x - \alpha u_{3x} &=& 0.
\end{eqnarray}
Requiring uniformity in rank, we obtain (after some algebra) 
\begin{equation}
\label{boussinesqweights}
W(u) = 2, \quad W(v) = 3, \quad W(\beta) = 2, \quad
W(\frac{\partial}{\partial t}) = 2. 
\end{equation}
Therefore, (\ref{boussinesqsystem}) is invariant under the scaling symmetry
\begin{equation}
\label{boussinesqscale}
(x, t, u, v, \beta) \rightarrow 
(\lambda^{-1} x, \lambda^{-2} t, \lambda^2 u, \lambda^3 v, \lambda^2 \beta). 
\end{equation}
\vskip 2pt
\noindent
{\bf Definition}:
System (\ref{continuoussystem}) is called {\it multi-uniform} in rank
if it admits more than one scaling symmetry 
(none of which result from introducing unnecessary parameters with weights).
\vskip 3pt
\noindent
{\bf Example}:
The LL system (\ref{heisenberg}) is not uniform in rank unless we allow 
the parameters $\alpha, \beta,$ and $\gamma$ to have weights.
Doing so, uniformity in rank requires
\begin{eqnarray}
\label{weightsheisenberg}
\!\!\!\!\!\!\!W(u) \!+\! W(\frac{\partial}{\partial t}) 
\!\!\!&\!=\!&\!\!\! W(v) \!+\! W(w) \!+\! 2 
\!=\! W(\gamma) \!+\! W(v) \!+\! W(w) 
\!=\! W(\beta) \!+\! W(v) \!+\! W(w), 
\nonumber \\
\!\!\!\!\!\!\!W(v) \!+\! W(\frac{\partial}{\partial t}) 
\!\!\!&\!=\!&\!\!\! W(u) \!+\! W(w) \!+\! 2 
\!=\! W(\alpha) \!+\! W(u) \!+\! W(w) 
\!=\! W(\gamma) \!+\! W(u) \!+\! W(w), 
\\
\!\!\!\!\!\!\!W(w) \!+\! W(\frac{\partial}{\partial t}) 
\!\!\!&\!=\!&\!\!\! W(u) \!+\! W(v) \!+\! 2 
\!=\! W(\beta) \!+\! W(u) \!+\! W(v) 
\!=\! W(\alpha) \!+\! W(u) \!+\! W(v). 
\nonumber 
\end{eqnarray}
Solving these relations gives  
$W(u) = W(v) = W(w), \; W(\alpha) = W(\beta) = W(\gamma) = 2, \; 
W(\frac{\partial}{\partial t}) = W(u) + 2, $ where $W(u)$ is arbitrary. 
The LL equation is thus multi-uniform and invariant under the following
class of scaling symmetries, 
\begin{equation}
\label{heisenbergcalegeneral}
(x, t, u, v, w, \alpha, \beta, \gamma) \rightarrow 
(\lambda^{-1} x, \lambda^{-(a+2)} t, \lambda^a u, \lambda^a v, \lambda^a w,
\lambda^2 \alpha, \lambda^2 \beta, \lambda^2 \gamma),
\end{equation}
where $W(u) = a$ is an arbitrary non-negative integer or rational number. 
For example, if we take $a = 1$ then
$W(u) = W(v) = W(w) = 1, \; W(\alpha) = W(\beta) = W(\gamma) = 2, \; 
W(\frac{\partial}{\partial t}) = 3$ and (\ref{heisenberg}) 
is invariant under 
%
$ 
(x, t, u, v, w, \alpha, \beta, \gamma) \rightarrow 
(\lambda^{-1} x, \lambda^{-3} t, \lambda u, \lambda v, \lambda w,
\lambda^2 \alpha, \lambda^2 \beta, \lambda^2 \gamma). 
$
%
Taking different choices for $a < W(\alpha) = 2$ will prove advantageous 
for the computations in Section~\ref{applll}.
\vskip 3pt
\noindent
{\bf Example}:
The SWW equations (\ref{sww}) are not uniform in rank unless we give a weight 
to $\Omega.$
Indeed, uniformity in rank for (\ref{sww}) requires, after some algebra, that
\begin{eqnarray}
\label{swwweightequations}
&& W({\partial/\partial t}) =
W(\Omega), \;
W({\partial/\partial y}) = W({\partial/\partial x}) = 1, 
\; W(u) = W(v) = W(\Omega) - 1, 
\nonumber \\
&& W(\theta) =
2 W(\Omega) - W(h) - 2, 
\end{eqnarray} 
where $W(h)$ and $W(\Omega)$ are arbitrary. 
Hence, the SWW system is multi-uniform and invariant under the class of 
scaling symmetries, 
\begin{equation}
\label{swwscaleclass}
(x, y, t, u, v, h, \theta, \Omega) \rightarrow 
(\lambda^{-1} x, \lambda^{-1} y, \lambda^{-b} t, 
\lambda^{b-1} u, \lambda^{b-1} v, 
\lambda^a h, \lambda^{2b-a-2} \theta, \lambda^b \Omega).
\end{equation}
where $W(h) \!=\! a$ and $W(\Omega) \!=\! b.$
Various choices for $a$ and $b$ will aid the computations in 
Section~\ref{applsww}.
\section{Conservation Laws}
\label{conservationlaws}
{\bf Definition}:
A {\it conservation law} \cite{PObook93} in differential form is a PDE 
of the form, 
\begin{equation} 
\label{pdeconslaw} 
{\rm D}_{t} \, \rho + {\rm Div} \, {\bf J} = 0, 
\end{equation}
which is satisfied on solutions of (\ref{continuoussystem}). 
The total time derivative ${\rm D}_t$ and total divergence ${\rm Div}$
are defined below.
The scalar differential function $\rho$ is called the 
{\it conserved density}; the vector differential function ${\bf J}$ 
is the associated {\it flux}.
In electromagnetism, (\ref{pdeconslaw}) is the continuity equation relating 
{\it charge density} $\rho$ to {\it current} ${\bf J}.$

In general, $\rho$ and ${\bf J}$ are functions of ${\bf x}, t, {\bf u},$ 
and the derivatives of ${\bf u}$ with respect to the components of ${\bf x}.$
In this paper, we will assume that densities and fluxes 
(i) are {\it local}, which means free of integral terms; 
(ii) {\it polynomial} in ${\bf u}$ and its derivatives; and
(iii) do not explicitly depend on ${\bf x}$ and $t.$ 
So, $\rho({\bf u}, {\bf u}_x, {\bf u}_y, {\bf u}_z, {\bf u}_{2x}, \dots)$
and ${\bf J}({\bf u}, {\bf u}_x, {\bf u}_y, {\bf u}_z, {\bf u}_{2x}, \dots).$
\vskip 3pt
\noindent
{\bf Definition}:
There is a close relationship between conservation laws and {\it constants 
of motion}.
Indeed, integration of (\ref{pdeconslaw}) over all space yields the 
integral form of a conservation law,
\begin{equation} 
\label{const}
P = 
\int_{-\infty}^{\infty} \int_{-\infty}^{\infty} \int_{-\infty}^{\infty} 
\rho \; dx\, dy\, dz = {\rm constant},
\end{equation}
provided that ${\bf J}$ vanishes at infinity.
As in mechanics, the $P\/$'s are called constants of motion.

The total divergence {\rm Div}
is computed as 
$ {\rm Div} \, {\bf J} 
= ( {\rm D}_x, {\rm D}_y, {\rm D}_z ) \cdot ( J_1, J_2, J_3 ) 
= {\rm D}_x J_1 + {\rm D}_y J_2 + {\rm D}_z J_3. $ 
In the 1D case, with one spatial variable $(x),$ conservation equation 
(\ref{pdeconslaw}) reduces to 
\begin{equation}
\label{1dpdeconslaw}
{\rm D}_t \, \rho + {\rm D}_x J = 0, 
\end{equation}
where both density $\rho$ and flux $J$ are scalar differential functions. 
\vskip 3pt
\noindent
The conservation laws (\ref{pdeconslaw}) and (\ref{1dpdeconslaw}) involve 
total derivatives ${\rm D}_t$ and ${\rm D}_x.$
In the 1D case, 
\begin{equation}
\label{operatordt}
{\rm D}_t \rho = \frac{\partial \rho}{\partial t} 
+ \sum_{k=0}^{n} \frac{\partial \rho}{\partial u_{kx}} {\rm D}_x^k (u_t).
\end{equation}
where $n$ is the order of $u$ in $\rho.$
Upon replacement of $u_t, u_{tx},$ etc.\ from $u_t = G,$ we get
\begin{equation}
\label{operatordtfrechet}
{\rm D}_t \rho = 
\frac{\partial \rho}{\partial t} + \rho(u)^{\prime}[G],
\end{equation}
where $\rho(u)^{\prime}[G]$ is the Fr\'echet derivative of $\rho$ in the 
direction of $G.$
Similarly, 
\begin{equation}
\label{operatordx}
{\rm D}_x J 
= \frac{\partial J}{\partial x} 
+ \sum_{k=0}^{m} \frac{\partial J}{\partial u_{kx}} u_{(k+1)x}, 
\end{equation}
where $m$ is the order of $u$ of $J.$
\vskip 3pt
\noindent
{\bf Example}:
Assume that the KdV equation (\ref{kdv}) has a density of the form 
\begin{equation}
\label{kdvrhoassumed}
\rho = c_1 u^3 + c_2 u_x^2, 
\end{equation}
where $c_1$ and $c_2$ are constants. 
Since $\rho$ does not explicitly depend on time and is of order $n=1$ in $u$, 
application of (\ref{operatordt}) gives
\begin{eqnarray}
\label{dtkdvrhoassumed}
{\rm D}_t \rho &=& \frac{\partial \rho}{\partial u} {\rm I} (u_t) 
+ \frac{\partial \rho}{\partial u_x} {\rm D}_x (u_t) 
\nonumber \\
&=& 3 c_1 u^2 u_t + 2 c_2 u_x u_{tx}  = 
 - 3 c_1 u^2 (u u_x + u_{3x}) - 2 c_2 u_x (u u_x + u_{3x})_x 
\nonumber \\
&=& - 3 c_1 u^2 (u u_x + u_{3x}) - 2 c_2 u_x (u_x^2 + u u_{2x} + u_{4x}) 
\nonumber \\
&=& -( 3 c_1 u^3 u_x + 3 c_1 u^2 u_{3x} + 2 c_2 u_x^3 
+ 2 c_2 u u_x u_{2x} + 2 c_2 u_x u_{4x} ), 
\end{eqnarray}
where ${\rm D}_x^0 = {\rm I}$ is the identity operator and where 
we used (\ref{kdv}) to replace $u_t$ and $u_{tx}.$
The generalizations of (\ref{operatordt}) and (\ref{operatordx}) to multiple 
dependent variables is straightforward \cite{PObook93}.
\vskip 3pt
\noindent
{\bf Example}:
Taking ${\bf u}(x,t) = (u(x,t),v(x,t),w(x,t)),$
\begin{eqnarray}
\label{operatordtforuandv}
{\rm D}_t \rho 
&=& \frac{\partial \rho}{\partial t} 
+ \sum_{k=0}^{n_1} \frac{\partial \rho}{\partial u_{kx}} {\rm D}_x^k (u_t) 
+ \sum_{k=0}^{n_2} \frac{\partial \rho}{\partial v_{kx}} {\rm D}_x^k (v_t)
+ \sum_{k=0}^{n_3} \frac{\partial \rho}{\partial w_{kx}} {\rm D}_x^k (w_t), 
\\
\label{operatordxforuandv}
{\rm D}_x J 
&=& \frac{\partial J}{\partial x} 
+ \sum_{k=0}^{m_1} \frac{\partial J}{\partial u_{kx}} u_{(k+1)x} 
+ \sum_{k=0}^{m_2} \frac{\partial J}{\partial v_{kx}} v_{(k+1)x}
+ \sum_{k=0}^{m_3} \frac{\partial J}{\partial w_{kx}} w_{(k+1)x}
\end{eqnarray}
where $n_1, n_2$ and $n_3$ are the highest orders of $u,v$ and $w$ in $\rho,$
and $m_1, m_2$ and $m_3$ are the highest orders of $u,v,w$ in $J.$
\vskip 4pt
\noindent
{\bf Example}:
Assume that the Boussinesq system (\ref{boussinesqsystem}) has a density 
of the form 
\begin{equation}
\label{boussinesqrhoassumed} 
\rho = 
c_1 \beta^2 u + c_2 \beta u^2 + c_3 u^3 + c_4 v^2 + c_5 u_x v + c_6 u_x^2. 
\end{equation}
where $c_1$ through $c_6$ are constants. 
Since $n_1 \!=\!1$ and $n_2 \!=\! 0$, application of 
(\ref{operatordtforuandv}) gives
\begin{eqnarray}
\label{dtboussinesqrhoassumed}
\!\!\!\!\!{\rm D}_t \rho \!&\!\!=\!\!&\! 
\frac{\partial \rho}{\partial u} {\rm I} (u_t)
+ \frac{\partial \rho}{\partial u_{x}} {\rm D}_x (u_t)
+ \frac{\partial \rho}{\partial v} {\rm I} (v_t)
\nonumber \\
\!\!&\!\!=\!\!&\! 
(c_1 \beta^2 + 2 c_2 \beta u + 3 c_3 u^2) u_t 
+ (c_5 v + 2 c_6 u_x) u_{tx} 
+ (2 c_4 v + c_5 u_x) v_t
\nonumber \\
\!\!&\!\!=\!\!&\! 
\!-\!\left( 
(c_1 \beta^2 \!+\! 2 c_2 \beta u \!+\! 3 c_3 u^2) v_x 
\!+\! (c_5 v \!+\! 2 c_6 u_x) v_{2x} 
\!+\! (2 c_4 v \!+\! c_5 u_x) (\beta u_x \!-\! 3 u u_x \!-\! \alpha u_{3x})
\right)\!,
\end{eqnarray}
where we replaced $u_t, u_{tx},$ and $v_t$ from (\ref{boussinesqsystem}).
\vskip 2pt
\noindent
{\bf Remark}:
The flux ${\bf J}$ in (\ref{pdeconslaw}) is not uniquely defined.
In 1D we use (\ref{1dpdeconslaw}) and the flux is only determined up 
to an arbitrary constant. 
In 2D, the flux is only determined up to a divergence-free vector 
${\bf K} = (K_1, K_2) = ( {\rm D}_y \phi, -{\rm D}_x \phi),$ 
where $\phi$ is an arbitrary scalar differential function. 
In 3D, the flux can only be determined up to a curl term. 
Indeed, if $( \rho, {\bf J} )$ is a valid density-flux pair, so is 
$( \rho, {\bf J} + {\mbox {\boldmath $\nabla$}} \times {\bf K} )$ for any 
arbitrary vector differential function ${\bf K} = (K_1, K_2, K_3).$ 
Recall that ${\mbox {\boldmath $\nabla$}} \times {\bf K} = 
({\rm D}_y K_3 - {\rm D}_z K_2, {\rm D}_z K_1 - {\rm D}_x K_3, 
{\rm D}_x K_2 - {\rm D}_y K_1).$ 
\vskip 4pt
\noindent
{\bf Example}: 
The Korteweg-de Vries equation (\ref{kdv}) is known 
\cite{MAandPCbook91,MAandHSbook81}
to have infinitely many polynomial conservation laws. 
The first three density-flux pairs are
\begin{eqnarray}
\label{kdvconslaws}
\rho^{(1)} \!&\!=\!&\! u, 
\quad\quad\quad\quad\quad\;\;
J^{(1)} = \frac{1}{2} u^2 + u_{2x}, 
\nonumber \\
\rho^{(2)} \!&\!=\!&\! \frac{1}{2} u^2, 
\quad \quad\quad\quad\;\;
J^{(2)} = \frac{1}{3} u^3 - \frac{1}{2} u_x^2 + u u_{2x}, 
\\
\rho^{(3)} \!&\!=\!&\! \frac{1}{3} u^3 - u_x^2, 
\quad\quad\;\,
J^{(3)} =  
\frac{1}{4} u^4 - 2 u u_x^2 + u^2 u_{2x} + u_{2x}^2 - 2 u_x u_{3x}.
\nonumber 
\end{eqnarray}
The first two express conservation of momentum and energy, respectively.
They are easy to compute by hand. 
The third one, less obvious and requiring more work, corresponds to 
Boussinesq's moment of instability \cite{AN83}.
Observe that the above densities are uniform of ranks $2, 4,$ and $6.$
The fluxes are also uniform of ranks $4, 6,$ and $8.$

The computations get quickly out of hand for higher ranks.
For example, for rank $12$
\begin{equation}
\rho^{(6)} = \frac{1}{6} u^6 - 10 u^3 u_x^2 - 5 u_x^4 + 18 u^2 {u_{2x}}^2 
+ \frac{120}{7} {u_{2x}}^3 - \frac{108}{7} u {u_{3x}}^2 
+ \frac{36}{7} {u_{4x}}^2 
\end{equation}
and $J^{(6)}$ is a scaling homogeneous polynomial with $20$ terms of 
rank 14 (not shown).
\vskip 3pt
\noindent
In general, if in (\ref{1dpdeconslaw}) ${\rm rank}(\rho) = R$ 
then ${\rm rank}(J) = R + W({\partial/\partial t}) - 1 .$ 
All the pieces in (\ref{pdeconslaw}) are also uniform in rank.
This comes as no surprise since the conservation law (\ref{pdeconslaw})
holds on solutions of (\ref{continuoussystem}), hence it `inherits' 
the dilation symmetry of (\ref{continuoussystem}).
\vskip 4pt
\noindent
{\bf Example}:
The Boussinesq equation (\ref{boussinesq}) also admits infinitely many 
conservation laws and is completely integrable 
\cite{MAandPCbook91,MAandHSbook81}.
The first four density-flux pairs \cite{PAthesis03} for 
(\ref{boussinesqsystem}) are 
\begin{eqnarray}
\label{boussinesqconslaws}
\rho^{(1)} \!&\!=\!&\! u, 
\quad\quad\quad\quad\quad\quad\quad\quad\quad\quad\;\,
J^{(1)} = v , 
\nonumber \\
\rho^{(2)} \!&\!=\!&\! v, 
\quad\quad\quad\quad\quad\quad\quad\quad\quad\quad\;\,
J^{(2)} = \beta u - \frac{3}{2} u^2 - \alpha u_{2x},
\\ 
\rho^{(3)} \!&\!=\!&\! u v, 
\quad\quad\quad\quad\quad\quad\quad\quad\quad\;\;\;
J^{(3)} = \frac{1}{2} \beta u^2 - u^3 + \frac{1}{2}v^2 
+ \frac{1}{2} \alpha u_x^2 - \alpha u u_{2x},
\nonumber \\
\rho^{(4)} \!&\!=\!&\! \beta u^2 - u^3 + v^2 + \alpha u_x^2, 
\quad\quad
J^{(4)} = 2 \beta u v - 3 u^2 v - 2 \alpha u_{2x} v + 2 \alpha u_x v_x .
\nonumber 
\end{eqnarray}
The densities are of ranks $2,3,5$ and $6,$ respectively.
The corresponding fluxes are of one rank higher. 
After setting $\beta = 1$ we obtain the conserved quantities of 
(\ref{boussinesqsys}) even though initially this system was not uniform 
in rank.
\vskip 4pt
\noindent
{\bf Example}: 
We assume that $\alpha, \beta,$ and $\gamma$ in (\ref{landaulifshitz}) are 
nonzero.
Cases with vanishing parameters must be investigated separately.
The first six density-flux pairs \cite{LFandLT87,MSandHF93} are
\begin{eqnarray}
\label{heisenbergconslaws}
\rho^{(1)} \!&\!=\!&\! u, 
\quad\quad\quad\quad\quad\quad\quad\quad\quad\;
J^{(1)} = v_x w - v w_x, \quad (\beta = \gamma \ne \alpha), 
\nonumber \\
\rho^{(2)} \!&\!=\!&\! v, 
\quad \quad\quad\quad\quad\quad\quad\quad\quad\;
J^{(2)} = u w_x - u_x w, \quad (\alpha = \gamma \ne \beta), 
\nonumber \\
\rho^{(3)} \!&\!=\!&\! w, 
\quad\quad\quad\quad\quad\quad\quad\quad\quad
J^{(3)} = u_x v - u v_x, \quad\;\; (\alpha = \beta \ne \gamma), 
\nonumber \\
\rho^{(4)} \!&\!=\!&\! u^2 + v^2 + w^2, 
\quad\quad\quad\quad\;
J^{(4)} = 0, 
\\
\rho^{(5)} \!&\!=\!&\! (u^2 + v^2 + w^2)^2, 
\quad\quad\quad 
J^{(5)} = 0, 
\nonumber \\
\rho^{(6)} \!&\!=\!&\! u_x^2 + v_x^2 + w_x^2 + 
(\gamma - \alpha) u^2 + (\gamma - \beta) v^2, 
\nonumber \\ 
J^{(6)} \!&\!=\!&\! 
2 \Big( 
(v w_x - v_x w) u_{2x} + (u_x w - u w_x) v_{2x} + (u v_x - u_x v) w_{2x}
\Big.
\nonumber \\
&& \Big. + (\beta - \gamma) u_x v w + (\gamma - \alpha) u v_x w
     + (\alpha - \beta) u v w_x \Big). 
\nonumber 
\end{eqnarray}
Note that the second and third density-flux pairs follow from the first pair 
via the cyclic permutations in (\ref{cyclicheisenberg}).
If we select $W(u) = W(v) = W(w) = a = \frac{1}{4},$ 
then the first three densities have rank $\frac{1}{4}$ and 
$\rho^{(4)}, \rho^{(5)},$ and $\rho^{(6)}$ have ranks 
$\frac{1}{2},1,\frac{5}{2},$ respectively.

The physics of (\ref{landaulifshitz}) demand that the magnitude 
$S=||{\bf S}||$ of the spin field is constant in time.
Indeed, using (\ref{heisenberg}) we readily verify that 
$u u_t + v v_t + w w_t = 0.$ 
Hence, any differentiable function of $S$ is also conserved in time.
This is apparent in $\rho^{(4)} = ||{\bf S}||^2$ and
$\rho^{(5)} = ||{\bf S}||^4,$ which are conserved (in time) even for
the fully anisotropic case where $\alpha \ne \beta \ne \gamma \ne \alpha.$ 

Obviously, linear combinations of conserved densities are also conserved. 
Hence, 
$\rho^{(6)} = {||{\bf S}_x||}^2 + (\gamma-\alpha) u^2 + (\gamma-\beta) v^2 $ 
can be replaced by 
\begin{equation}
\label{heisenberghamiltonian1}
{\tilde{\rho}}^{(6)} = \frac{1}{2} ( \rho^{(6)} - \gamma \rho^{(4)} ) 
=  \frac{1}{2} ( u_x^2 + v_x^2 + w_x^2 - \alpha u^2 - \beta v^2 - \gamma w^2).
\end{equation}
Now, $u_x^2 = {\rm D}_x (u u_x) - u u_{2x},$ etc., and densities are 
divergence-equivalent (see Section~\ref{algorithm}) if they differ by a 
total derivative with respect to $x.$
So, (\ref{heisenberghamiltonian1}) is equivalent with 
\begin{equation}
\label{heisenberghamiltonian2}
{\tilde{\rho}}^{(6)} = 
- \frac{1}{2} 
( u u_{2x}  + v v_{2x} + w w_{2x} + \alpha u^2 + \beta v^2 + \gamma w^2), 
\end{equation}
which can be compactly written as 
\begin{equation}
\label{heisenberghamiltonian3}
{\tilde{\rho}}^{(6)} = 
- \frac{1}{2} ( {\bf S} \cdot \Delta {\bf S} + {\bf S} \cdot D {\bf S} ),
\end{equation}
where $\Delta$ is the Laplacian and $D = {\rm diag}(\alpha,\beta,\gamma).$ 
So, ${\tilde{\rho}}^{(6)}$ is the Hamiltonian density \cite{MSandHF93} in 
\begin{eqnarray}
\label{heisenberghamiltonian}
H &=& - \frac{1}{2} \int \left( {\bf S} \cdot \Delta {\bf S} + 
      {\bf S} \cdot D {\bf S} \right) \; dx 
\nonumber \\
& = & \frac{1}{2} \int \left(
{||{\bf S}_x||}^2 + (\gamma-\alpha) u^2 + (\gamma-\beta) v^2 \right) \; dx .
\end{eqnarray}
The Hamiltonian $H$ expresses that the total energy is constant in time 
since $dH/dt = 0.$
\vskip 4pt
\noindent
{\bf Example}:
The SWW equations (\ref{sww}) admit a Hamiltonian formulation \cite{PDpf03}
and infinitely many conservation laws.
Yet, that is still insufficient to guarantee complete integrability of PDEs 
in $(2+1)$-dimensions. 
The first few conserved densities and fluxes for (\ref{sww}) are 
\begin{equation}
\begin{array}{llll}
\label{swwconslaws}
\hspace*{-1.2mm} \rho^{(1)} = h,  & 
{\bf J}^{(1)} =
\left(
\begin{array}{c}
 u h  \\
 v h  
\end{array}
\right), \quad\quad
&
\rho^{(2)} = h \theta , 
&
{\bf J}^{(2)} =
\left(\!
\begin{array}{c}
 u h \theta \\
 v h \theta 
\end{array}
\!\right), 
\\[3mm]
\hspace*{-1.2mm} \rho^{(3)} =h \theta^2 ,  & 
{\bf J}^{(3)} =
\left(\!
\begin{array}{c}
 u h \theta^2  \\
 v h \theta^2 
 \end{array}
 \!\right), & &
\\[3mm]
\multicolumn{2}{l}{\hspace*{-1.2mm} \rho^{(4)} = (u^2 + v^2) h + h^2 \theta ,} 
&
\multicolumn{2}{l}{{\bf J}^{(4)} =
\left(\!
\begin{array}{c}
 u^3 h + u v^2 h + 2 u h^2 \theta \\
 v^3 h + u^2 v h + 2 v h^2 \theta 
\end{array}
\!\right),}
\\[3mm]
\multicolumn{4}{l}{\hspace*{-1.2mm} \rho^{(5)} =
(2 \Omega - u_y + v_x) \theta,}
\\[3mm]
\multicolumn{4}{l}{\hspace*{-1.2mm} {\bf J}^{(5)} =
\frac{1}{6} \left(
\begin{array}{c}
12 \Omega u \theta - 4 u u_y \theta + 6 u v_x \theta + 2 v v_y \theta 
+ u^2 \theta_y + v^2 \theta_y - h \theta \theta_y + h_y \theta^2 
\\
12 \Omega v \theta + 4 v v_x \theta - 6 v u_y \theta - 2 u u_x \theta 
- u^2 \theta_x -\! v^2 \theta_x + h \theta \theta_x - h_x \theta^2 
\end{array}
\right).}
\end{array}
\end{equation}
As shown in \cite{PDpf03}, system (\ref{sww}) has conserved densities 
$\rho = h f(\theta)$ and $\rho = (2 \Omega - u_y + v_x) g(\theta),$ 
where $f(\theta)$ and $g(\theta)$ are arbitrary functions. 
Such densities are the integrands of the Casimirs of the Poisson bracket 
associated with (\ref{sww}). 
The algorithm presented in Section~\ref{applsww} only computes densities of 
the form $\rho = h \theta^k$ and $\rho = (2 \Omega - u_y + v_x) \theta^l,$
where $k$ and $l$ are positive integers. 
\section{Tools from the Calculus of Variations}
\label{tools}
In this section we introduce the variational derivative (Euler operator),
the higher Euler operators (also called Lie-Euler operators) from the 
calculus of variations, and the homotopy operator from homological 
algebra and variational bi-complexes \cite{IAbook04}.
These tools will be applied in Section~\ref{applications}.
\subsection{Variational Derivative (Euler Operator)}
\label{zeroeuler}
{\bf Definition}:
A scalar differential function $f$ is a {\it divergence} if and only if  
there exists a vector differential function ${\bf F}$ such that 
$f = {\rm Div} \, {\bf F}.$ 
In 1D, we say that a differential function $f$ is 
{\it exact}\footnote{We do not use {\it integrable} to avoid confusion with
complete integrability from soliton theory 
\cite{MAandPCbook91,MAandHSbook81}.} 
if and only if there exists a scalar differential function $F$ such that 
$f = {\rm D}_x F.$ 
Obviously, $F = {\rm D}_x^{-1}(f) = \int f \, dx$ is then the primitive 
(or integral) of $f.$
\vskip 3pt
\noindent
{\bf Example}:
Taking $f = -{\rm D}_t \rho$ from (\ref{dtkdvrhoassumed}), that is 
\begin{equation}
\label{fcontinuous}
f = 3 c_1 u^3 u_x + 2 c_2 u_x^3 + 2 c_2 u u_x u_{2x} 
     + 3 c_1 u^2 u_{3x} + 2 c_2 u_x u_{4x}, 
\end{equation}
where\footnote{Variable $t$ is parameter in subsequent computations.
For brevity, we write $u(x)$ instead of $u(x;t).$} 
$u(x;t)$, we will compute $c_1$ and $c_2$ so that $f$ is exact.

Upon repeated integration by parts (by hand), we get
\begin{equation}
\label{Fcontinuousbyhand}
F = \int f \, dx = 
\frac{3}{4} c_1 u^4 + (c_2 - 3 c_1) u u_x^2 + 3 c_1 u^2 u_{2x} 
    - c_2 u_{2x}^2 + 2 c_2 u_x u_{3x} + \int (3 c_1 + c_2) u_x^3 \; dx .
\end{equation}
Removing the ``obstruction" (the last term) requires $c_2 \!=\! -3 c_1.$ 
Substituting a solution, $c_1 \!=\! \frac{1}{3}, c_2 \!=\! -1,$ 
into (\ref{kdvrhoassumed}) gives $\rho^{(3)}$ in (\ref{kdvconslaws}).
Substituting that solution into (\ref{fcontinuous}) and 
(\ref{Fcontinuousbyhand}) yields 
\begin{equation}
\label{fcontinuousexact}
f = u^3 u_x - 2 u_x^3 - 2 u u_x u_{2x} + u^2 u_{3x} - 2 u_x u_{4x} 
\end{equation}
which is exact, together with its integral 
\begin{equation}
\label{Fcontinuousexact}
F = \frac{1}{4} u^4 - 2 u u_x^2 + u^2 u_{2x} + u_{2x}^2 - 2 u_x u_{3x},
\end{equation}
which is $J^{(3)}$ in (\ref{kdvconslaws}).
Currently, CAS like {\it Mathematica}, 
{\it Maple}\footnote{Future versions of {\it Maple} will be able to 
``partially" integrate such expressions \cite{BDwebsite04,WHetalbook05}.},
and {\it REDUCE}  cannot compute (\ref{Fcontinuousbyhand}) as a sum of terms 
that can be integrated out and the obstruction.
\vskip 4pt
\noindent
{\bf Example}:
Consider the following example in 2D 
\begin{equation}
\label{divergenceuv}
f = u_x v_y - u_{2x} v_y - u_y v_x + u_{xy} v_x, 
\end{equation}
where $u(x,y)$ and $v(x,y).$
It is easy to verify by hand (via integration by parts) that $f$ is exact. 
Indeed, $f = {\rm Div} \, {\bf F}$ with 
\begin{equation}
\label{vectorfordivergenceuv}
{\bf F} = (u v_y - u_x v_y, -u v_x + u_x v_x).
\end{equation}
\vskip 3pt
\noindent
As far as we know, the leading CAS currently lack tools to compute ${\bf F}.$
Two questions arise: 
\vskip 2pt
\noindent
(i) How can we tell whether $f$ is the divergence of some differential 
function ${\bf F}?$
\vskip 2pt
\noindent
(ii) How do we compute ${\bf F}$ automatically and without multivariate 
integration by parts of expressions involving arbitrary functions?
\vskip 2pt
\noindent
To answer these questions we use the following tools from the calculus of 
variations: the variational derivative (Euler operator), its generalizations 
(higher Euler operators or Lie-Euler operators), and the homotopy operator.
%
\noindent
\vskip 5pt
\noindent
{\bf Definition}:
The {\it variational derivative} (zeroth Euler operator), 
${\cal {L}}^{({\bf 0})}_{ {\bf u}({\bf x})}$, is defined 
\cite[p.\ 246]{PObook93} by 
\begin{equation}
\label{zeroeulervectoruvectorx}
{\cal {L}}^{({\bf 0})}_{{\bf u}({\bf x})} = 
\sum_J (-{\rm D})_J \frac{\partial}{\partial {\bf u}_J}, 
\end{equation}
where the sum is over all the unordered multi-indices $J$ 
\cite[p.\ 95]{PObook93}.
For example, in the 2D case the multi-indices corresponding to second-order 
derivatives can be identified with $\{2x,2y,2z,xy,xz,yz\}.$
Obviously, 
$(-D)_{2x} = (-{\rm D}_x)^2 $$ = {\rm D}_x^2 $, 
$(-D)_{xy}$ $ = (-{\rm D}_x) (-{\rm D}_y) $ $ = {\rm D}_x {\rm D}_y,$ 
etc..
For notational details see \cite[p.\ 95, p.\ 108, p.\ 246]{PObook93}.

With applications in mind, we give calculus-based formulas for the 
variational derivatives in 1D and 2D.
The formula for 3D is analogous and can be found in \cite{WHetalbook05}.
\vskip 3pt
\noindent
{\bf Example}:
In 1D, where ${\bf x} = x,$ for scalar component $u(x)$ 
we have\footnote{Variable $t$ in ${\bf u}({\bf x};t)$ is suppressed 
because it is a parameter in the Euler operators.}
\begin{equation}
\label{zeroeulerscalarux}
{\cal {L}}^{(0)}_{u(x)} \!=\!
\sum_{k=0}^{\infty} (-{\rm D}_x)^k \frac{\partial}{\partial u_{kx} } 
\!=\! \frac{\partial}{\partial {u} } 
- {\rm D}_x \frac{\partial}{\partial {u_x} }
+ {\rm D}_{x}^2 \frac{\partial }{\partial {u_{2x}} } 
- {\rm D}_{x}^3 \frac{\partial }{\partial {u_{3x}} } 
+ \cdots,  
\end{equation}
In 2D where ${\bf x} = (x,y),$ we have for component $u(x,y)$ 
\begin{eqnarray}
\label{zeroeulerscalaruxy}
{\cal {L}}^{(0,0)}_{u(x,y)} 
\!\!&\!=\!&\!\!
\sum_{k_{x}=0}^{\infty} \; \sum_{k_{y}=0}^{\infty}
(-{\rm D}_x)^{k_{x}} (-{\rm D}_y)^{k_{y}} 
\frac{\partial }{\partial u_{k_{x}x \, k_{y}y}}
\!=\! \frac{\partial}{\partial {u} } 
-\! {\rm D}_x \frac{\partial}{\partial {u_x} }
-\! {\rm D}_y \frac{\partial}{\partial {u_y} }
\nonumber \\
&&\!\!+\, {\rm D}_{x}^2 \frac{\partial }{\partial {u_{2x}} } 
+\! {\rm D}_{x} {\rm D}_{y} \frac{\partial }{\partial {u_{xy}} } 
+\, {\rm D}_{y}^2 \frac{\partial }{\partial {u_{2y}} } 
-\, {\rm D}_{x}^3 \frac{\partial }{\partial{ u_{3x}} } 
- \cdots ,
\end{eqnarray}
Note that ${u}_{k_{x}x \, k_{y}y}$ stands for 
${u}_{xx \cdots x \, yy \cdots y}$ where $x$ is repeated $k_x$ times 
and $y$ is repeated $k_y$ times. 
Similar formulas hold for components $v,w,$ etc..
The first question is then answered by the following theorem 
\cite[p.\ 248]{PObook93}.
\vskip 5pt
\noindent
{\bf Theorem}:
A necessary and sufficient condition for a function $f$ 
to be a divergence, i.e.\ there exists a differential function ${\bf F}$ 
so that $f = {\rm Div} \, {\bf F},$ is that 
${\cal {L}}_{{\bf u}({\bf x})}^{({\bf 0})} (f) \equiv 0.$
In words: the Euler operator annihilates divergences.
\vskip 4pt
\noindent 
If, for example, ${\bf u}({\bf x}) = (u({\bf x}),v({\bf x}))$ then both 
${\cal {L}}_{u({\bf x})}^{({\bf 0})} (f)$ 
and ${\cal {L}}_{v({\bf x})}^{({\bf 0})} (f)$ must vanish identically.
For the 1D case, the theorem says that a differential function $f$ is exact, 
i.e.\ there exists a differential function $F$ so that $f = D_x F,$ 
if and only if ${\cal {L}}^{(0)}_{{\bf u}(x)} (f) \equiv 0.$
\vskip 3pt
\noindent
{\bf Example}: 
Avoiding integration by parts, we determine $c_1$ and $c_2$ so that 
$f$ in (\ref{fcontinuous}) will be exact. 
Since $f$ is of order $4$ in the (one) dependent variable $u(x),$ 
the zeroth Euler operator (\ref{zeroeulerscalarux}) terminates at $k=4.$
Using nothing but differentiations, we readily compute
\begin{eqnarray}
\label{computationzeroeulerux}
{\cal {L}}^{(0)}_{u(x)} (f) 
\!&\!=\!&\!
\frac{\partial f}{\partial u} 
- {\rm D}_x \left( \frac{\partial f}{\partial u_x} \right)
+ {\rm D}_{x}^2 \left( \frac{\partial f}{\partial u_{2x}} \right)
- {\rm D}_{x}^3 \left( \frac{\partial f}{\partial u_{3x}} \right)
+ {\rm D}_{x}^4 \left( \frac{\partial f}{\partial u_{4x}} \right)
\nonumber \\
\!&\!=\!&\! 
9 c_1 u^2 u_x + 2 c_2 u_x u_{2x} + 6 c_1 u u_{3x}
- {\rm D}_x ( 3 c_1 u^3 + 6 c_2 u_x^2 + 2 c_2 u u_{2x} + 2 c_2 u_{4x} )
\nonumber \\
&& + {\rm D}_x^2 ( 2 c_2 u u_x ) - {\rm D}_x^3 ( 3 c_1 u^2 ) + 
{\rm D}_x^4 ( 2 c_2 u_x )
\nonumber \\
\!&\!=\!&\! 9 c_1 u^2 u_x + 2 c_2 u_x u_{2x} + 6 c_1 u u_{3x} 
- (9 c_1 u^2 u_x + 14 c_2 u_x u_{2x} + 2 c_2 u u_{3x} + 2 c_2 u_{5x})
\nonumber \\
&& + ( 6 c_2 u_x u_{2x} + 2 c_2 u u_{3x} )
- ( 18 c_1 u_x u_{2x} + 6 c_1 u u_{3x} )
\nonumber \\
\!&\!=\!&\! - 6 (3 c_1 + c_2) u_x u_{2x}.
\end{eqnarray}
Note that the terms in $u^2 u_x, u u_{3x},$ and $u_{5x}$ dropped out. 
Hence, ${\cal {L}}^{(0)}_{u(x)} (f) \equiv 0$ leads to
$3 c_1 \!+\! c_2 = 0.$
Substituting $c_1 \!=\! \frac{1}{3}, c_2 \!=\! -1$ into (\ref{fcontinuous}) 
gives (\ref{fcontinuousexact}) as claimed.
\vskip 3pt
\noindent
{\bf Example}:
As an example in 2D, we readily verify that 
$f =  u_x v_y - u_{2x} v_y - u_y v_x + u_{xy} v_x$ 
from (\ref{divergenceuv}) is a divergence. 
Applying (\ref{zeroeulerscalaruxy}) to $f$ for each component of 
${\bf u}(x,y) = (u(x,y),v(x,y))$ separately, we straightforwardly verify that 
${\cal {L}}^{(0,0)}_{u(x,y)} (f) \equiv 0$ and 
${\cal {L}}^{(0,0)}_{v(x,y)} (f) \equiv 0.$
\subsection{Removing Divergences and Divergence-equivalent Terms}
\label{algorithm}
It is of paramount importance that densities are free of divergences for 
they could be moved into the flux ${\bf J}$ of the conservation law 
${\rm D}_{t}\,\rho + {\rm Div}\,{\bf J} = 0.$
We show how the Euler operator can be used to remove divergences and 
divergence-equivalent terms in densities.
An algorithm to do so is given in \cite{WHetalbook05}.
The following examples illustrate the concept behind the algorithm. 
\vskip 4pt
\noindent
{\bf Definition}:
Two scalar differential functions, $f^{(1)}$ and $f^{(2)},$ are 
{\it divergence-equivalent} if and only if they differ by the divergence 
of some vector ${\bf V},$ i.e.\ $f^{(1)} \sim f^{(2)}$ if and only if 
$f^{(1)} - f^{(2)} = {\rm Div} \, {\bf V}.$
If a scalar expression is divergence-equivalent to zero then it is a 
divergence.
\vskip 4pt
\noindent
{\bf Example}:
Any conserved density of the KdV equation (\ref{kdv}) is uniform 
in rank with respect to $W(u) \!=\! 2$ and $W({\partial/\partial x}) \!=\! 1.$ 
The list of all terms of, say, rank 6 is
$\!{\cal {R}} \!=\! \{ u^3, u_x^2, u u_{2x}, u_{4x} \}.$  
Now, terms $f^{(1)} = u u_{2x}$ and $f^{(2)} = - u_x^2$ are 
divergence-equivalent because 
$f^{(1)} - f^{(2)} = u_x^2 + u u_{2x} = {\rm D}_x (u u_x).$
Using (\ref{zeroeulerscalarux}), note that 
${\cal {L}}_{u(x)}^{(0)} (-u_x^2) = 2 u_{2x}$ and 
${\cal {L}}_{u(x)}^{(0)} (u u_{2x}) = 2 u_{2x}$ are equal
(therefore, linearly dependent). 
So, we discard $u u_{2x}.$ 
Moreover, $u_{4x} = {\rm D}_x (u_{3x})$ is a divergence and, 
as expected, ${\cal {L}}_{u(x)}^{(0)} (u_{4x}) = 0.$ 
So, we can discard $u_{4x}.$
Hence, ${\cal {R}} $ can be replaced by ${\cal {S}} \!=\! \{ u^3, u_x^2 \}$ 
which is free of divergences and divergence-equivalent terms.
\vskip 3pt
\noindent
{\bf Example}:
The list of non-constant terms of rank 6 for the Boussinesq equation 
(\ref{boussinesqsystem}) is
\begin{equation}
\label{boussinesqsetr}
{\cal {R}} = \{ \beta^2 u, \beta u^2, u^3, v^2, u_x v, u_x^2,
\beta v_x, u v_x, \beta u_{2x}, u u_{2x}, v_{3x}, u_{4x} \}.
\end{equation}
Using (\ref{zeroeulerscalarux}), for every term $t_i$ in ${\cal {R}}$ 
we compute ${\bf v}_i = {\cal {L}}_{{\bf u}(x)}^{(0)} (t_i) = 
({\cal {L}}_{u(x)}^{(0)} (t_i), {\cal {L}}_{v(x)}^{(0)} (t_i)).$
If ${\bf v}_i = (0,0)$ then $t_i$ is discarded and so is ${\bf v}_i.$
If ${\bf v}_i \ne (0,0)$ we verify whether or not ${\bf v}_i$ is linearly 
independent of the non-zero vectors ${\bf v}_j,\,$ $j=1,2, \cdots, i-1.$ 
If independent, the term $t_i$ is kept, otherwise, $t_i$ is discarded and so
is ${\bf v}_i.$

Upon application of (\ref{zeroeulerscalarux}), 
the first six terms in ${\cal {R}}$ lead to linearly independent vectors 
${\bf v}_1$ through ${\bf v}_6.$
Therefore, $t_1$ through $t_6$ are kept 
(and so are the corresponding vectors). 
For $t_7 = \beta v_x$ we compute 
${\bf v}_7 \!=\! {\cal {L}}_{{\bf u}(x)}^{(0)} (\beta v_x) \!=\!(0,0).$ 
So, $t_7$ is discarded and so is ${\bf v}_7.$
For $t_8 = u v_x$ we get 
${\bf v}_8 \!=\! {\cal {L}}_{{\bf u}(x)}^{(0)} (u v_x)
\!=\!(v_x, -u_x) = -{\bf v}_5.$ 
So, $t_8$ is discarded and so is ${\bf v}_8.$

Proceeding in a similar fashion, $t_{9}, t_{10}, t_{11}$ and $t_{12}$ 
are discarded.
Thus, ${\cal {R}}$ is replaced by 
\begin{equation}
\label{boussinesqsets}
{\cal {S}} \!=\! \{ \beta^2 u, \beta u^2, u^3, v^2, u_x v, u_x^2 \},
\end{equation}
which is free of divergences and divergence-equivalent terms.
\vskip 3pt
\noindent
{\bf Example}:
In the 2D case, $f^{(1)} = (u_x - u_{2x}) v_y$ and 
$f^{(2)} = (u_y - u_{xy} ) v_x $ are divergence-equivalent since  
$f^{(1)} - f^{(2)} = u_x v_y - u_{2x} v_y - u_y v_x + u_{xy} v_x  = 
{\rm Div} \, ( u v_y - u_x v_y , - u v_x + u_x v_x).$
Using (\ref{zeroeulerscalaruxy}), note that
${\cal {L}}_{{\bf u}(x,y)}^{(0)} (f^{(1)}) =
{\cal {L}}_{{\bf u}(x,y)}^{(0)} (f^{(2)}) = 
(-v_{xy} - v_{xxy}, - u_{xy} + u_{xxy}).$
\subsection{Higher Euler Operators}
\label{highereuler}
To compute ${\bf F} = {\rm Div}^{-1}(f)$ or, in the 1D case,  
$F = {\rm D}_x^{-1}(f) = \int f \,dx,$ we need higher-order versions of the
variational derivative, called {\it higher Euler operators} 
\cite{MKetal70,PObook93}
or {\it Lie-Euler operators} \cite{IAbook04}.
The general formulas in terms of differential forms are given in 
\cite[p.\ 367]{PObook93}. 
We give calculus-based formulas for the 1D and 2D cases 
(see \cite{WHetalbook05} for the 3D case).
\vskip 4pt
\noindent
{\bf Definition}:
The {\it higher Euler operators} in 1D 
(with variable $x$)\footnote{Variable $t$ in ${\bf u}({\bf x};t)$ 
is suppressed because it is a parameter in the higher Euler operators.} are  
\begin{equation}
\label{highereulervectorux}
{\cal {L}}^{(i)}_{{\bf u}(x)} = 
\sum_{k=i}^{\infty} {k \choose i} (-{\rm D}_x)^{k-i} 
\frac{\partial }{\partial {\bf u}_{kx}}, 
\end{equation}
where ${k \choose i}$ is the binomial coefficient.
Note that the higher Euler operator for $i=0$ matches the variational 
derivative in (\ref{zeroeulerscalarux}).
\vskip 3pt
\noindent
{\bf Example}:
The first three higher Euler operators in 1D for component $u(x)$ are
\begin{eqnarray}
\label{highereulerscalarux}
{\cal {L}}^{(1)}_{u(x)} 
\!\!&\!=\!&\!\!
\frac{\partial }{\partial u_x} 
- 2{\rm D}_x     \frac{\partial }{\partial u_{2x}}
+ 3{\rm D}_{x}^2 \frac{\partial }{\partial u_{3x}} 
- 4{\rm D}_{x}^3 \frac{\partial }{\partial u_{4x}} + \cdots ,
\nonumber \\
%
{\cal {L}}^{(2)}_{u(x)} 
\!\!&\!=\!&\!\!
\frac{\partial }{\partial u_{2x}} 
- 3{\rm D}_x \frac{\partial }{\partial u_{3x}}
+ 6{\rm D}_{x}^2 \frac{\partial }{\partial u_{4x}}
- 10{\rm D}_{x}^3 \frac{\partial }{\partial u_{5x}} + \cdots ,
\\
{\cal {L}}^{(3)}_{u(x)} 
\!&\!=\!&\!
\frac{\partial }{\partial u_{3x}} 
- 4{\rm D}_x \frac{\partial }{\partial u_{4x}}
+ 10{\rm D}_{x}^2 \frac{\partial }{\partial u_{5x}}
- 20{\rm D}_{x}^3 \frac{\partial }{\partial u_{6x}} 
+ \cdots .
\nonumber
\end{eqnarray}
\vskip 3pt
\noindent
{\bf Definition}:
The {\it higher Euler operators} in 2D (with variables $x,y$) are 
\begin{equation}
\label{highereulervectoruxy}
{\cal {L}}^{(i_{x},i_{y})}_{{\bf u}(x,y)} =
\sum_{k_{x}=i_{x}}^{\infty} \sum_{k_{y}=i_{y}}^{\infty}
{k_{x} \choose i_{x}} {k_{y} \choose i_{y}} 
(-{\rm D}_x)^{k_{x}-i_{x}}  (-{\rm D}_y)^{k_{y}-i_{y}} 
\frac{\partial }{\partial {\bf u}_{k_{x}x \, k_{y}y}} .
\end{equation}
\noindent
Note that the higher Euler operator for $i_{x}\!=\!i_{y}\!=\!0$ matches 
the variational derivative in (\ref{zeroeulerscalaruxy}).
\vskip 3pt
\noindent
{\bf Example}:
The first higher Euler operators for component $u(x,y)$ are 
\begin{eqnarray}
\label{highereulerscalaruxy}
{\cal {L}}^{(1,0)}_{u(x,y)} 
\!&\!=\!&\!
\frac{\partial }{\partial u_x} 
-2 {\rm D}_x  \frac{\partial }{\partial u_{2x}}
-  {\rm D}_y  \frac{\partial }{\partial u_{xy}}
+ 3 {\rm D}_{x}^2  \frac{\partial }{\partial u_{3x}} 
+ 2 {\rm D}_{x} {\rm D}_{y} \frac{\partial }{\partial u_{2xy}} 
-\cdots,
\nonumber \\
{\cal {L}}^{(0,1)}_{u(x,y)} 
\!&\!=\!&\!
\frac{\partial }{\partial u_y} 
- 2{\rm D}_y     \frac{\partial }{\partial u_{2y}}
- {\rm D}_x    \frac{\partial }{\partial u_{yx}}
+ 3 {\rm D}_{y}^2 \frac{\partial }{\partial u_{3y}} 
+ 2 {\rm D}_{x} {\rm D}_{y} \frac{\partial }{\partial u_{x2y}} 
-\cdots, 
\\
{\cal {L}}^{(1,1)}_{u(x,y)} 
\!&\!=\!&\!
\frac{\partial }{\partial u_{xy}} 
- 2 {\rm D}_x \frac{\partial }{\partial u_{2xy}}
- 2 {\rm D}_{y} \frac{\partial }{\partial u_{x2y}}
+ 3 {\rm D}_{x}^2 \frac{\partial }{\partial u_{3xy}} 
+ 4 {\rm D}_{x} {\rm D}_{y} \frac{\partial }{\partial u_{2x2y}} 
+ \cdots,
\nonumber \\
{\cal {L}}^{(2,1)}_{u(x,y)} 
\!&\!=\!&\!
\frac{\partial }{\partial u_{2xy}} 
- 3 {\rm D}_x \frac{\partial }{\partial u_{3xy}}
-2 {\rm D}_{y} \frac{\partial }{\partial u_{2x2y}}
+ 6 {\rm D}_{x}^2 \frac{\partial }{\partial u_{4xy}} 
+ {\rm D}_{y}^2 \frac{\partial }{\partial u_{2x3y}} 
- \cdots .
\nonumber
\end{eqnarray}
The higher Euler operators are useful in their own right as the following 
theorem \cite{MKetal70} in 1D shows. 
\vskip 4pt
\noindent
{\bf Theorem}:
A necessary and sufficient condition for a function $f$ 
to be an $r^{\rm th}$ order derivative, i.e.\ there exists a scalar 
differential function $F$ so that 
$f = D_x^r F,$ is that ${\cal {L}}^{(i)}_{{\bf u}(x)} (f) \equiv 0$ for 
$i\!=\!0,1,\ldots,r\!-\!1.$
\subsection{Homotopy Operators}
\label{homotopy}
Armed with the higher Euler operators we now turn to the homotopy operator 
which will allow us to reduce the computation of 
${\bf F} = {\rm Div}^{-1}(f),$ 
or in the 1D case 
$F = {\rm D}_x^{-1}(f) = \int f \, dx,$
to a single integral with respect to an auxiliary variable 
denoted by $\lambda$ (not to be confused with $\lambda$ in 
Section~\ref{dilationinvariance}).
Amazingly, the homotopy operator circumvents integration by parts 
(in multi-dimensions involving arbitrary functions) and reduces the 
inversion of the total divergence operator, ${\rm Div},$ 
to a problem of single-variable calculus. 

As mentioned in Section~\ref{zeroeuler}, ${\rm Div}^{-1}$ is defined 
up to a divergence-free (curl) term.
In 3D, ${\rm Div}^{-1}$ is an equivalence class ${\rm Div}^{-1}(f) = 
{\bf F} + {\mbox {\boldmath $\nabla$}} \times {\bf K}$ 
where ${\bf K}$ is an arbitrary vector differential function.
The homotopy operator computes a particular ${\bf K}.$

The homotopy operator in terms of differential forms is given in 
\cite[p.\ 372]{PObook93}.
Below we give calculus-based formulas for the homotopy operators in 
1D and 2D which are easy to implement in CAS.
The explicit formulas in 3D are analogous (see \cite{WHetalbook05}).
\vskip 3pt
\noindent
{\bf Definition}:
The {\it homotopy operator} in 1D 
(with variable $x$)\footnote{Variable $t$ in ${\bf u}({\bf x};t)$ 
is suppressed because it is a parameter in the homotopy operators.} is
\begin{equation}
\label{homotopyvectorux}
{\cal {H}}_{{\bf u}(x)}(f) = 
\int_{0}^{1} \sum_{j=1}^{N} I_{u_j}(f) [\lambda {\bf u}] 
\, \frac{d \lambda}{\lambda},
\end{equation}
where $u_j$ is the $j$th component of ${\bf u}$ and the integrand $I_{u_j}(f)$ 
is given by 
\begin{equation}
\label{integrandhomotopyvectorux}
I_{u_j}(f) = \sum_{i=0}^{\infty} {\rm D}_{x}^i 
\left( u_j \, {\cal {L}}^{(i+1)}_{u_j(x)} (f) \right).
\end{equation}
The integrand involves the 1D higher Euler operators in 
(\ref{highereulervectorux}).
In (\ref{homotopyvectorux}), $N$ is the number of dependent variables and 
$I_{u_j}(f)[\lambda {\bf u}]$ means that in 
$I_{u_j}(f)$ we replace ${\bf u}(x) \rightarrow \lambda {\bf u}(x), \, 
{\bf u}_x(x) \rightarrow \lambda {\bf u}_x(x),\,$ etc..
In practice, we first add the $I_{u_j}(f)$ and then scale the variables.
The sum in $I_{u_j}(f)$ terminates at $i=p-1$ where $p$ is the order 
of $u_j$ in $f.$ 
 
Given an exact function $f,$ the second question, that is how to compute 
$F = {\rm D}_x^{-1}(f) = \int f \, dx,$ is then answered by the following 
theorem \cite[p.\ 372]{PObook93}.
\vskip 5pt
\noindent
{\bf Theorem}:
For an exact function $f,$ one has $F = {\cal {H}}_{{\bf u}(x)}(f).$
\vskip 4pt
\noindent 
Thus, in the 1D case, applying the homotopy operator 
(\ref{homotopyvectorux}) allows one to bypass integration by parts.
As an experiment, one can start from some function ${\tilde F},$ 
compute $f = {\rm D}_x {\tilde F},$ then compute 
$F = {\cal {H}}_{{\bf u}(x)}(f),$ and finally verify that 
$F - {\tilde F}$ is a constant.
\vskip 3pt
\noindent
{\bf Example}:
Using the homotopy operator (\ref{homotopyvectorux}) with 
(\ref{integrandhomotopyvectorux}), we recompute (\ref{Fcontinuousexact}).
Since $f$ in (\ref{fcontinuous})
%
%
is of order $p=4$ in $u_1(x) = u(x),$ the sum in 
(\ref{integrandhomotopyvectorux}) terminates at $i = p-1 = 3.$
Hence, 
\begin{eqnarray} 
\label{Iuforf}
I_{u}(f) \!\!&\!=\!&\!\! 
                     u {\cal {L}}^{(1)}_{u(x)} (f) 
+ {\rm D}_x \left(   u {\cal {L}}^{(2)}_{u(x)} (f) \right) 
+ {\rm D}_x^2 \left( u {\cal {L}}^{(3)}_{u(x)} (f) \right) 
+ {\rm D}_x^3 \left( u {\cal {L}}^{(4)}_{u(x)} (f) \right) 
\nonumber \\
\!\!&\!=\!&\!\!   u \frac{\partial f}{\partial u_x}
  - 2 u {\rm D}_x   \left(\! \frac{\partial f}{\partial u_{2x}} \!\right)
  + 3 u {\rm D}_x^2 \left(\! \frac{\partial f}{\partial u_{3x}} \!\right)
  - 4 u {\rm D}_x^3 \left(\! \frac{\partial f}{\partial u_{4x}} \!\right) 
+ {\rm D}_x \left(\!
         u \frac{\partial f}{\partial u_{2x}} 
     - 3 u {\rm D}_x   \left(\! \frac{\partial f}{\partial u_{3x}} \!\right)
\right.
\nonumber \\
&& \left. 
+ 6 u {\rm D}_x^2 \left(\! \frac{\partial f}{\partial u_{4x}} \!\right)
     \!\right) 
+ {\rm D}_x^2 \left(\!
         u \frac{\partial f}{\partial u_{3x}} 
     - 4 u {\rm D}_x \left(\! \frac{\partial f}{\partial u_{4x}} \!\right)
     \!\right) 
   + {\rm D}_x^3 \left(\!
         u \frac{\partial f}{\partial u_{4x}} 
     \!\right)  
\nonumber \\
\!\!&\!=\!&\!\!
u^4 + 8 u^2 u_{2x} + 6 u u_{4x} + 4 u u_x^2
- 4 {\rm D}_x \left( 2 u^2 u_x + 3 u u_{3x} \right)
+ {\rm D}_x^2 \left( u^3 + 8 u u_{2x} \right)
- 2 {\rm D}_x^3 \left(u u_x \right)
\nonumber \\ 
\!\!&\!=\!&\!\!
u^4 - 6 u u_x^2 + 3 u^2 u_{2x} + 2 u_{2x}^2 - 4 u_x u_{3x}.
\end{eqnarray}
Formula (\ref{homotopyvectorux}), with ${\bf u} = u_1 = u,$ 
requires an integration with respect to $\lambda:$
\begin{eqnarray}
\label{F}
F \!\!&\!\!=\!\!&\!\! {\cal {H}}_{u(x)} (f) 
= \int_0^1\! I_{u} (f)[\lambda u] \,\frac{d\lambda}{\lambda} 
\nonumber \\ 
\!\!&\!\!=\!\!&\!\!\int_0^1\!\left( 
\lambda^3 u^4 - 6 \lambda^2 u u_x^2 + 3 \lambda^2 u^2 u_{2x} 
+ 2 \lambda u_{2x}^2 - 4 \lambda u_x u_{3x} \right) d\lambda  
\nonumber \\ 
\!\!&\!\!=\!\!&\!\! 
\frac{1}{4} u^4 - 2 u u_x^2 + u^2 u_{2x} + u_{2x}^2 - 2 u_x u_{3x}.
\end{eqnarray}
\vskip 1pt
\noindent
The crux of the homotopy operator method is that the integration by parts 
of a differential expression like (\ref{fcontinuousexact}), which involves 
an arbitrary function $u(x)$ and its derivatives, can be reduced to a 
standard integration of a polynomial in $\lambda.$
\vskip 3pt
\noindent
{\bf Example}:
For a system with $N=2$ components, 
${\bf u}(x) = (u_1(x),u_2(x)) = (u(x),v(x)),$ the homotopy operator formulas 
are
\begin{equation}
\label{homotopyuvx}
{\cal {H}}_{{\bf u}(x)} (f) = 
\int_{0}^{1} \left( I_{u}(f) + I_{v}(f) \right) [\lambda {\bf u}]
\, \frac{d \lambda}{\lambda},
\end{equation}
with
\begin{equation}
\label{integrandhomotopyuvx}
I_{u}(f) = \sum_{i=0}^{\infty} {\rm D}_{x}^i 
\left( u \, {\cal {L}}^{(i+1)}_{u(x)} (f) \right)
\quad {\rm and} \quad
I_{v}(f) = \sum_{i=0}^{\infty} {\rm D}_{x}^i 
\left( v \, {\cal {L}}^{(i+1)}_{v(x)} (f) \right).
\end{equation}
\vskip 3pt
\noindent
{\bf Example}: 
Consider 
$f = 3 u_x v^2 \sin u - u_x^3 \sin u - 6 v v_x \cos u + 2 u_x u_{2x} \cos u 
+ 8 v_x v_{2x}, $ 
which is no longer polynomial in the two $(N=2)$ dependent variables 
$u(x)$ and $v(x).$ 
Applying Euler operator (\ref{zeroeulervectoruvectorx}) for each 
component of ${\bf u}(x) = (u(x),v(x))$ separately, we quickly verify that 
${\cal {L}}^{(0)}_{u(x)} (f) \equiv 0$ 
and ${\cal {L}}^{(0)}_{v(x)} (f) \equiv 0.$
Hence, $f$ is exact. 
Integration by parts (by hand) gives 
\begin{equation}
\label{Fcontinuousbyhandsincos}
F = \int f \, dx = 4 v_x^2 + u_x^2 \cos u -3 v^2 \cos u.
\end{equation}
Currently, CAS like {\it Mathematica}, 
{\it Maple}\footnote{Version 9.5 of {\it Maple} can integrate 
such expressions as a result of our interactions with the developers.} 
and {\it Reduce} fail this integration due to the presence of trigonometric 
functions.

We now recompute (\ref{Fcontinuousbyhandsincos}) with the homotopy 
operator formulas (\ref{homotopyuvx}) and (\ref{integrandhomotopyuvx}).
First,
\begin{eqnarray} 
\label{Iuforfsincos}
I_{u}(f) &=& 
u {\cal {L}}^{(1)}_{u(x)} (f) + 
{\rm D}_x \left( u {\cal {L}}^{(2)}_{u(x)} (f) \right) 
= u \frac{\partial f}{\partial u_x} 
    - 2 u {\rm D}_x \left(\! \frac{\partial f}{\partial u_{2x}} \!\right)
    + {\rm D}_x \left(\! u \frac{\partial f}{\partial u_{2x}} \!\right) 
\nonumber \\
&=& 3 u v^2 \sin u - u u_x^2 \sin u + 2 u_x^2 \cos u.
\end{eqnarray} 
\noindent 
Second, 
\begin{eqnarray} 
\label{Ivforfsincos}
I_{v}(f) &=& 
v {\cal {L}}^{(1)}_{v(x)} (f) + 
{\rm D}_x \left(\! v {\cal {L}}^{(2)}_{v(x)} (f) \!\right) 
= v \frac{\partial f}{\partial v_x} 
    - 2 v {\rm D}_x \left(\! \frac{\partial f}{\partial v_{2x}} \!\right)
    + {\rm D}_x \left(\! v \frac{\partial f}{\partial v_{2x}} \!\right) 
\nonumber \\
&=& - 6 v^2 \cos u + 8 v_x^2.
\end{eqnarray} 
Finally, using (\ref{homotopyuvx}), 
\begin{eqnarray}
\label{Fsincos}
F \!\!&\!\!=\!\!&\!\! {\cal {H}}_{{\bf u}(x)} (f) 
= \int_0^1\!\left( I_{u} (f) + I_{v} (f) \right) [\lambda {\bf u}] 
\, \frac{d\lambda}{\lambda} 
\nonumber \\ 
\!\!&\!\!=\!\!&\!\!\int_0^1\!\Big( 3 \lambda^2 u v^2 \sin(\lambda u) 
  \!-\! \lambda^2 u u_x^2 \sin(\lambda u) 
  \!+\!2 \lambda u_x^2 \cos(\lambda u) 
\!-\! 6 \lambda v^2 \cos(\lambda u) \!+\!8 \lambda v_x^2 
\Big) d\lambda  
\nonumber \\ 
\!\!&\!\!=\!\!&\!\! u_x^2 \cos u - 3 v^2 \cos u + 4 v_x^2.
\end{eqnarray}
In contrast to the integration in (\ref{Fcontinuousbyhandsincos}) which 
involved arbitrary functions $u(x)$ and $v(x),$ in (\ref{Fsincos}) 
we integrated by parts with respect to the auxiliary variable $\lambda.$
Any CAS can do that! 
\vskip 4pt
\noindent
We now turn to inverting the ${\rm Div}$ operator using the homotopy operator.
\vskip 4pt
\noindent
{\bf Definition}:
We define the {\it homotopy operator} in 2D (with variables $x,y$)  
through its two components $( {\cal {H}}^{(x)}_{{\bf u}(x,y)}(f), 
$$ {\cal {H}}^{(y)}_{{\bf u}(x,y)}(f) ).$
The $x$-component of the operator is given by
\begin{equation}
\label{homotopyvectoruxycompx}
{\cal {H}}^{(x)}_{{\bf u}(x,y)} (f) = 
\int_{0}^{1} \sum_{j=1}^{N} I_{u_j}^{(x)} (f)
[\lambda {\bf u}] \, \frac{d \lambda}{\lambda},
\end{equation}
with
\begin{equation}
\label{integrandhomotopyvectoruxycompx}
I_{u_j}^{(x)}(f) = 
\sum_{i_x=0}^{\infty} \sum_{i_y=0}^{\infty} 
\left( \frac{1+i_x}{1+i_x+i_y} \right)
{\rm D}_{x}^{i_x} {\rm D}_{y}^{i_y} 
\left( u_j \, {\cal {L}}^{(1+i_x,i_y)}_{u_j(x,y)} (f) \right).
\end{equation}
Likewise, the $y$-component is given by
\begin{equation}
\label{homotopyvectoruxycompy}
{\cal {H}}^{(y)}_{{\bf u}(x,y)} (f) = 
\int_{0}^{1} \sum_{j=1}^{N} I_{u_j}^{(y)} (f) 
[\lambda {\bf u}] \, \frac{d \lambda}{\lambda},
\end{equation}
with 
\begin{equation}
\label{integrandhomotopyvectoruxycompy}
I_{u_j}^{(y)} (f) = 
\sum_{i_x=0}^{\infty} \sum_{i_y=0}^{\infty} 
\left( \frac{1+i_y}{1+i_x+i_y} \right)
{\rm D}_{x}^{i_x} {\rm D}_{y}^{i_y}
\left( u_j \, {\cal {L}}^{(i_x,1+i_y)}_{u_j(x,y)} (f) \right).
\end{equation}
Integrands (\ref{integrandhomotopyvectoruxycompx}) and 
(\ref{integrandhomotopyvectoruxycompy}) involve the 2D higher Euler 
operators in (\ref{highereulervectoruxy}).
The question how to compute ${\bf F} = (F_1,F_2) = {\rm Div}^{-1}(f)$ 
is then answered by the following theorem.
\vskip 4pt
\noindent
{\bf Theorem}:
If $f$ is a divergence, then
$ {\bf F} = (F_1, F_2) = {\rm Div}^{-1}(f) = 
({\cal {H}}^{(x)}_{{\bf u}(x,y)} (f), {\cal {H}}^{(y)}_{{\bf u}(x,y)} (f) ). $ 
\vskip 2pt
\noindent 
The superscripts $(x)$ and $(y)$ remind us which components of ${\bf F}$ 
we are computing.
As a matter of testing, we can start from some vector ${\tilde{\bf F}}$ 
and compute $f = {\rm Div} \, {\tilde {\bf F}}.$ 
Next, we compute ${\bf F} = (F_1, F_2) = 
({\cal {H}}^{(x)}_{{\bf u}(x,y)} (f), 
{\cal {H}}^{(y)}_{{\bf u}(x,y)} (f) )$
and, finally, verify that 
${\bf K} = {\tilde{\mathbf F}} - {\mathbf F}$ is divergence free.
\vskip 4pt
\noindent
{\bf Example}:
Using (\ref{divergenceuv}), we show how application of the 2D homotopy 
operator leads to (\ref{vectorfordivergenceuv}), up to a divergence 
free vector. 
Consider $f = u_x v_y - u_{2x} v_y - u_y v_x + u_{xy} v_x$, 
which is the divergence of ${\bf F}$ in (\ref{vectorfordivergenceuv}).
In order to compute ${\rm Div}^{-1}(f)$, we use 
(\ref{integrandhomotopyvectoruxycompx}) for the $u$ component in 
${\bf u}(x,y) = (u(x,y), v(x,y)):$
\begin{eqnarray}
\label{Iuxf2D}
I_u^{(x)}(f) & = & u {\cal {L}}_{u(x,y)}^{(1,0)}(f)
+ {\rm D}_x \left( u{\cal {L}}_{u(x,y)}^{(2,0)}(f) \right)
+ \frac{1}{2}{\rm D}_y \left( u{\cal {L}}_{u(x,y)}^{(1,1)}(f) \right)
\nonumber \\
&=& u \left(\frac{\partial f}{\partial u_x}
- 2{\rm D}_x \frac{\partial f}{\partial u_{2x}}
- {\rm D}_y \frac{\partial f}{\partial u_{xy}}\right)
+ {\rm D}_x \left(u \frac{\partial f}{\partial u_{2x}} \right)
+ \frac{1}{2}{\rm D}_y \left( u \frac{\partial f}{\partial u_{xy}} \right)
\nonumber \\
&=& u v_y + \frac{1}{2} u_y v_x - u_x v_y + \frac{1}{2} u v_{xy}. 
\end{eqnarray}
Similarly, for the $v$ component of ${\bf u}(x,y) = (u(x,y),v(x,y))$ we get
\begin{equation}
\label{Ivxf2D}
I_v^{(x)}(f)= v {\cal {L}}_{v(x,y)}^{(1,0)}(f)
= v \frac{\partial f}{\partial v_x} = - u_y v + u_{xy} v. 
\end{equation}
Hence, using (\ref{homotopyvectoruxycompx}), 
\begin{eqnarray}
\label{applhomotop2Dxpart}
\!\!\!\!\! F_1 &\!\!=\!\!& {\cal {H}}^{(x)}_{{\bf u}(x,y)} (f) 
= \int_{0}^{1} 
\left( I_{u}^{(x)} (f) + I_{v}^{(x)} (f) \right) [\lambda {\bf u}] 
\, \frac{d \lambda}{\lambda}
\nonumber \\
&\!\!=\!\!& \int_0^1\!\lambda \left( 
u v_y + \frac{1}{2} u_y v_x - u_x v_y + \frac{1}{2} u v_{xy} 
- u_y v + u_{xy} v \right) \, d\lambda  
\nonumber \\ 
&\!\!=\!\!& 
\frac{1}{2} u v_y + \frac{1}{4} u_y v_x - \frac{1}{2} u_x v_y 
+ \frac{1}{4} u v_{xy}  - \frac{1}{2} u_y v + \frac{1}{2} u_{xy} v.
\end{eqnarray}
Without showing the details, using (\ref{homotopyvectoruxycompy}) and 
(\ref{integrandhomotopyvectoruxycompy}) we compute analogously 
\begin{eqnarray}
\label{applhomotop2Dypart}
\!\!\!\!\! F_2 &\!\!=\!\!& {\cal {H}}^{(y)}_{{\bf u}(x,y)} (f) 
= \int_{0}^{1} \left( I_{u}^{(y)} (f) + I_{v}^{(y)} (f) \right) 
[\lambda {\bf u}] \, \frac{d \lambda}{\lambda}
\nonumber \\
&\!\!=\!\!& \int_0^1\! \lambda \left(
- u v_x - \frac{1}{2} u v_{2x} + \frac{1}{2} u_x v_x 
+ u_x v - u_{2x} v \right) \, d\lambda  
\nonumber \\ 
&\!\!=\!\!& 
- \frac{1}{2} u v_x - \frac{1}{4} u v_{2x} + \frac{1}{4} u_x v_x 
+ \frac{1}{2} u_x v- \frac{1}{2} u_{2x} v.
\end{eqnarray}
Now we can readily verify that the resulting vector 
\begin{equation}
\label{vectorf2d}
{\bf F} = \left(
\begin{array}{c}
 F_1 \\
 F_2 \end{array} \right) = 
\left(\!
\begin{array}{c}
\frac{1}{2} u v_y + \frac{1}{4} u_y v_x - \frac{1}{2} u_x v_y 
+ \frac{1}{4} u v_{xy}  - \frac{1}{2} u_y v + \frac{1}{2} u_{xy} v \\
- \frac{1}{2} u v_x - \frac{1}{4} u v_{2x} + \frac{1}{4} u_x v_x 
+ \frac{1}{2} u_x v - \frac{1}{2} u_{2x} v \end{array}
\! \right)
\end{equation}
differs from ${\tilde{\bf F}} = (u v_y - u_x v_y , -u v_x + u_x v_x)$ by a 
divergence-free vector 
\begin{equation}
\label{vectorf2d-diff}
{\bf K} \!=\! {\tilde{\mathbf F}} - {\mathbf F} \!=\! 
\left(
\begin{array}{c}
 K_1 \\
 K_2 \end{array} \right) = 
\left(\!
\begin{array}{c}
\frac{1}{2} u v_y - \frac{1}{4} u_y v_x - \frac{1}{2} u_x v_y 
- \frac{1}{4} u v_{xy}  = \frac{1}{2} u_y v - \frac{1}{2} u_{xy} v \\
- \frac{1}{2} u v_x + \frac{1}{4} u v_{2x} + \frac{3}{4} u_x v_x 
- \frac{1}{2} u_x v + \frac{1}{2} u_{2x} v 
\end{array}
\! \right).
\end{equation}
As mentioned in Section~\ref{zeroeuler}, ${\bf K}$ can be written as 
$({\rm D}_y \phi, -{\rm D}_x \phi )$ with 
$\phi = \frac{1}{2} u v - \frac{1}{4} u v_x - \frac{1}{2} u_x v.$
\section{Computation of Conservation Laws}
\label{applications}
In this section we apply the Euler and homotopy operators to compute 
density-flux pairs for the four PDEs in Section~\ref{examples}.
Using the KdV equation (\ref{kdv}) we illustrate the steps for a 1D case 
(space variable $x$) and dependent variable $u(x,t).$
Part of the computations for this example were done in the previous sections.
The Boussinesq equation (\ref{boussinesqsystem}), also in 1D, is used to
illustrate the case with two dependent variables $u(x,t)$ and $v(x,t)$ and 
an auxiliary parameter with weight.

The LL equation (\ref{heisenberg}), still in 1D, has three dependent 
variables $u(x,t), v(x,t)$ and $w(x,t).$
The three coupling constants have weights which makes the computations more 
cumbersome. 
However, the multi-uniformity of (\ref{heisenberg}) helps to reduce the 
complexity. 
The shallow-water wave equations (\ref{sww}) illustrate the computations 
in 2D (two space variables) and four dependent variables. 
Again, we can take advantage of the multi-uniformity of (\ref{sww}) to
control ``expression swell" of the computations.

We use a direct approach to compute conservation laws, 
${\rm D}_{t}\,\rho + {\rm Div}\,{\bf J} = 0,$ of polynomial systems of 
nonlinear PDEs.
First, we build the candidate density $\rho$ as a linear combination 
(with constant coefficients $c_i)$ of terms that are uniform in rank 
with respect to the scaling symmetry of the PDE. 
In doing so, we dynamically remove divergences and divergence-equivalent 
terms to get the shortest possible candidate density. 

Second, we evaluate ${\rm D}_t \rho$ on solutions of the PDE, thus removing 
all time derivatives from the problem.
The resulting expression, $E = - {\rm D}_t \rho,$ must be a divergence 
of the as yet unknown flux. 
Thus, we compute ${\cal {L}}_{{\bf u}({\bf x})}^{({\bf 0})} (E)$ and
set the coefficients of like terms to zero.
This leads to a linear system for the undetermined coefficients $c_i.$
If the given PDE has arbitrary constant parameters, then the linear system
is parametrized. 
Careful analysis of the eliminant is needed to find all solution branches,
and, when applicable, the conditions on the parameters.
For each branch, the solution of the linear system is substituted into
$\rho$ and $E.$ 

Third, we use the homotopy operator ${\cal {H}}_{{\bf u}({\bf x})}$ 
to compute ${\bf J} \!=\!{\rm Div}^{-1}(E).$
The computations are done with our {\it Mathematica} packages 
\cite{WHwebsite04}.
Recall that ${\bf J}$ is only defined up a curl term.
Removing the curl term in ${\bf J}$ may lead to a shorter flux.
Inversion of ${\rm Div}$ via the homotopy operator therefore does not 
guarantee the shortest flux. 
\subsection{Conservation Laws for the KdV Equation}
\label{applkdv} 
In (\ref{kdvconslaws}) we gave the first three density-flux pairs of
(\ref{kdv}).
As an example, we will compute ${\rho}^{(3)}$ and $J^{(3)}.$ 
The weights are $W({\partial}/{\partial x}) = 1$ and $W(u) = 2.$ 
Hence, density ${\rho}^{(3)}$ has rank 6 and flux $J^{(3)}$ has rank 8.
The algorithm has three steps:
\vskip 2pt
\noindent
{\bf Step 1}: {\bf Construct the form of the density}
\vskip 2pt
\noindent
Start from ${\cal {V}} = \{ u \},$ i.e.\ the list of dependent variables 
(and parameters with weight, if applicable).
Construct the set ${\cal {M}}$ of all non-constant monomials of 
(selected) rank $6$ or less (without derivatives). 
Thus, ${\cal {M}} \!=\! \{ u^3, u^2, u \}$.
Next, for each monomial in ${\cal {M}},$ introduce the needed $x$-derivatives 
so that each term has rank 6.
Since $W({\partial}/{\partial x}) = 1,$ use
\begin{equation}
\label{buildingblockskdv}
\frac{{\partial}^2 u^2}{\partial{x^2}} 
\!=\! (2 u u_x)_x \!=\! 2 u_x^2 + 2 u u_{2x}, \quad\;
\frac{{\partial}^4 u}{\partial{x^4}} = u_{4x}.
\end{equation}
Ignore the highest-order terms (typically the last terms) in each 
of the right hand sides of (\ref{buildingblockskdv}). 
Augment ${\cal {M}}$ with the remaining terms 
(deleting numerical factors) to get ${\cal {R}} = \{ u^3, u_x^2 \}.$ 

Here ${\cal {S}} = {\cal {R}}$ since ${\cal {R}}$ is free of divergences 
and divergent-equivalent terms.
Linearly combine the terms in ${\cal {S}}$ with constant coefficients to 
get the candidate density:
\begin{equation}
\label{candidaterhokdv}
\rho = c_1 u^3 + c_2 u_x^2,
\end{equation}
which is (\ref{kdvrhoassumed}).
\vskip 2pt
\noindent
{\bf Step 2}: {\bf Compute the undetermined constants $c_i$}
\vskip 2pt
\noindent
Compute ${\rm D}_t \rho$ and use (\ref{kdv}) to eliminate $u_t$ and $u_{tx}.$ 
As shown in (\ref{dtkdvrhoassumed}), this gives
\begin{equation}
\label{Ekdv}
E = -{\rm D}_t \rho =
 3 c_1 u^3 u_x + 3 c_1 u^2 u_{3x} + 2 c_2 u_x^3 + 2 c_2 u u_x u_{2x} + 
 2 c_2 u_x u_{4x}.
\end{equation}
Since $E \!=\! -{\rm D}_t \, \rho \!=\! {\rm D}_x J,$ the expression $E$ 
must be exact.
As shown in (\ref{computationzeroeulerux}), 
${\cal {L}}^{(0)}_{u(x)} (E) \!=\! - 6 (3 c_1 + c_2) u_x u_{2x} \equiv 0$ 
leads to $c_1 \!=\! \frac{1}{3}, c_2 \!=\! -1$ 
and upon substitution into (\ref{candidaterhokdv}) to 
\begin{equation}
\label{rho3kdv}
\rho = \frac{1}{3} u^3 - u_x^2 
\end{equation}
which is $\rho^{(3)}$ in (\ref{kdvconslaws}).
\vskip 3pt
\noindent
{\bf Step 3}: {\bf Compute the flux $J$}
\vskip 2pt
\noindent
Substitute $c_1 \!=\! \frac{1}{3}, c_2 \!=\! -1$ into (\ref{Ekdv}) to get
\begin{equation}
\label{Ekdvsimp}
E = u^3 u_x - 2 u_x^3 - 2 u u_x u_{2x} + u^2 u_{3x} - 2 u_x u_{4x}.
\end{equation}
As shown in (\ref{F}), application of (\ref{homotopyvectorux}) with 
(\ref{integrandhomotopyvectorux}) to (\ref{Ekdvsimp}) gives
\begin{equation}
\label{j3kdv}
J = \frac{1}{4} u^4 - 2 u u_x^2 + u^2 u_{2x} + u_{2x}^2 - 2 u_x u_{3x}, 
\end{equation}
which is $J^{(3)}$ in (\ref{kdvconslaws}).
\subsection{Conservation Laws for the Boussinesq Equation}
\label{applboussinesq} 
The first few density-flux pairs of (\ref{boussinesqsystem}) were given 
in (\ref{boussinesqconslaws}). 
Equation (\ref{boussinesqsystem}) has weights
$W(\frac{\partial}{\partial x}) \!=\! 1$, $W(u) \!=\! W(\beta) \!=\! 2,$ 
and $W(v) \!=\! 3.$
We show the computation of ${\rho}^{(4)}$ and $J^{(4)}$ 
of ranks 6 and 7, respectively.
The presence of the auxiliary parameter $\beta$ with 
weight complicates matters.
\vskip 1pt
\noindent
{\bf Step 1}: {\bf Construct the form of the density}
\vskip 1pt
\noindent
Augment the list of dependent variables with the parameter $\beta$
(with non-zero weight). 
Hence, ${\cal {V}} = \{ u, v, \beta \}.$ 
Construct
${\cal {M}} = 
\{ \beta^2 u, \beta u^2, \beta u, \beta v, u^3, u^2, u, v^2, v, u v \},$
which contains all non-constant monomials of (chosen) rank $6$ or less 
(without derivatives). 
Next, for each term in ${\cal {M}},$ introduce the right number of 
$x$-derivatives so that each term has rank 6.
For example, 
\begin{equation}
\label{buildingblocksbous}
\frac{{\partial}^2 \beta u}{\partial{x^2}} \!=\! \beta u_{2x}, \quad
\frac{{\partial}^2 u^2}{\partial{x^2}} \!=\! 2 u_x^2 + 2 u u_{2x}, \quad
\frac{{\partial}^4 u}{\partial{x^4}} \!=\! u_{4x}, \quad
\frac{{\partial} ( u v )}{\partial{x}} = u_x v + u v_x, \quad etc..
\end{equation}
Ignore the highest-order terms (typically the last terms) in each of the 
right hand sides of (\ref{buildingblocksbous}). 
Augment ${\cal {M}} $ with the remaining terms, 
after deleting numerical factors, to get
${\cal {R}} = 
\{ \beta^2 u, \beta u^2, u^3, v^2, u_x v, u_x^2,
\beta v_x, u v_x, \beta u_{2x}, u u_{2x}, v_{3x}, u_{4x} \}$
as in (\ref{boussinesqsetr}).
As shown in (\ref{boussinesqsets}), 
removal of divergences and divergence-free terms in ${\cal {R}}$ leads to 
${\cal {S}} \!=\! \{ \beta^2 u, \beta u^2, u^3, v^2, u_x v, u_x^2 \}.$
Linearly combine the terms in ${\cal S}$ to get
\begin{equation}
\label{candidaterhoboussinesq}
\rho = 
c_1 \beta^2 u + c_2 \beta u^2 + c_3 u^3 + c_4 v^2 + c_5 u_x v + c_6 u_x^2. 
\end{equation}
\vskip 0.0001pt
\noindent
{\bf Step 2}: {\bf Compute the undetermined constants $c_i$}
\vskip 1pt
\noindent
Compute $E = -{\rm D}_t \rho$ and use (\ref{boussinesqsystem}) to eliminate 
$u_t, u_{tx},$ and $v_t.$ 
As shown in (\ref{dtboussinesqrhoassumed}), this gives
\begin{equation}
\label{Ebous}
E = (c_1 \beta^2 + 2 c_2 \beta u + 3 c_3 u^2) v_x 
+ (c_5 v + 2 c_6 u_x)  v_{2x} 
+ (2 c_4 v + c_5 u_x)  (\beta u_x - 3 u u_x - \alpha u_{3x}).
\end{equation}
$E \!=\! -{\rm D}_t \, \rho \!=\! {\rm D}_x J$ must be exact.
Thus, apply (\ref{zeroeulervectoruvectorx}) and 
require that ${\cal {L}}^{(0)}_{u(x)} (E) = {\cal {L}}^{(0)}_{v(x)} (E) 
\equiv 0.$ 
Group like terms. 
Set their coefficients equal to zero to obtain the parametrized system 
\begin{equation}
\label{systembous}
\beta (c_2 - c_4) = 0, \quad c_3 + c_4 = 0, \quad c_5 = 0, \quad 
\alpha c_5 = 0,\quad \beta c_5 = 0, \quad \alpha c_4 - c_6 = 0.
\end{equation}
Investigate the eliminant of the system. 
Set $c_1 = 1,$ to obtain the solution
\begin{equation}
\label{solcbous}
c_1 = 1, \quad c_2 = c_4, \quad 
c_3 = - c_4, \quad c_5 = 0, \quad c_6 = \alpha c_4,
\end{equation}
which is valid irrespective of the values of $\alpha$ and $\beta.$
Substitute (\ref{solcbous}) into (\ref{candidaterhoboussinesq}) to get
\begin{equation}
\label{rho4bousall}
\rho = \beta^2 u + c_4 ( \beta u^2 - u^3 + v^2 + \alpha u_x^2 ).
\end{equation}
The density must be split into independent pieces.
Indeed, since $c_4$ is arbitrary, set $c_4 = 0$ or $c_4 = 1,$ 
thus splitting (\ref{rho4bousall}) into two independent densities, 
$\rho = \beta^2 u$ and 
\begin{equation}
\label{rho4bous}
\rho = \beta u^2 - u^3 + v^2 + \alpha u_x^2,
\end{equation} 
which are $\rho^{(1)}$ and $\rho^{(4)}$ in (\ref{boussinesqconslaws}).
\vskip 3pt
\noindent
{\bf Step 3}: {\bf Compute the flux $J$}
\vskip 2pt
\noindent
Compute the flux corresponding to $\rho$ in (\ref{rho4bous}).
Substitute (\ref{solcbous}) into (\ref{Ebous}). 
Take the terms in $c_4$ and set $c_4 = 1.$
Thus, 
\begin{equation}
\label{Eboussimp}
E = 2 \beta u v_x + 2 \beta u_x v  - 3 u^2 v_x - 6 u u_x v 
+ 2 \alpha u_x v_{2x} - 2 \alpha u_{3x} v.
\end{equation}
Apply (\ref{homotopyuvx}) and (\ref{integrandhomotopyuvx}) to 
(\ref{Eboussimp}).
For $E$ of order $3$ in $u(x),$ compute
\begin{eqnarray} 
\label{IuforEbous}
I_{u}(E) \!&\!=\!&\! 
                     u {\cal {L}}^{(1)}_{u(x)} (E) 
+ {\rm D}_x \left(   u {\cal {L}}^{(2)}_{u(x)} (E) \right) 
+ {\rm D}_x^2 \left( u {\cal {L}}^{(3)}_{u(x)} (E) \right) 
\nonumber \\
\!&\!=\!&\!   u \frac{\partial E}{\partial u_x}
  - 2 u {\rm D}_x   \left(\! \frac{\partial E}{\partial u_{2x}} \!\right)
  + 3 u {\rm D}_x^2 \left(\! \frac{\partial E}{\partial u_{3x}} \!\right)
+ {\rm D}_x \left(\!
         u \frac{\partial E}{\partial u_{2x}} 
     - 3 u {\rm D}_x \left(\! \frac{\partial E}{\partial u_{3x}} \!\right)
     \!\right) 
+ {\rm D}_x^2 \left(\! 
         u \frac{\partial E}{\partial u_{3x}}
     \!\right) 
\nonumber \\
\!&\!=\!&\!
2 \beta u v - 6 u^2 v - 4 \alpha u v_{2x}
+ 6 \alpha {\rm D}_x \left( u v_x \right) 
- 2 \alpha {\rm D}_x^2 \left( u v \right)
\nonumber \\ 
\!\!&\!=\!&\!\!
2 \beta u v - 6 u^2 v - 2 \alpha u_{2x} v + 2 \alpha u_x v_x.
\end{eqnarray}
With $E$ is of order $2$ in $v(x),$ subsequently compute
\begin{eqnarray} 
\label{IvforEbous}
I_{v}(E) \!&\!=\!&\! 
                     v {\cal {L}}^{(1)}_{v(x)} (E) 
+ {\rm D}_x \left(   v {\cal {L}}^{(2)}_{v(x)} (E) \right) 
= v \frac{\partial E}{\partial v_x}
  - 2 v {\rm D}_x   \left(\! \frac{\partial E}{\partial v_{2x}} \!\right)
+ {\rm D}_x \left(\!
         v \frac{\partial E}{\partial v_{2x}} 
     \!\right) 
\nonumber \\
\!&\!=\!&\!
2 \beta u v - 3 u^2 v - 4 \alpha u v_{2x}
+ 2 \alpha {\rm D}_x \left( u_x v \right) 
= 2 \beta u v - 3 u^2 v - 2 \alpha u_{2x} v + 2 \alpha u_x v_x.
\end{eqnarray}
Formula (\ref{homotopyuvx}) with ${\bf u}(x) = (u(x),v(x))$ 
requires an integration with respect to $\lambda.$
Hence,
\begin{eqnarray}
\label{Fboussinesq}
J \!&\!=\!&\! {\cal {H}}_{u(x)} (E) 
= \int_0^1\! (I_{u}(E) + I_{v}(E))[\lambda {\bf u}] \,\frac{d\lambda}{\lambda} 
\nonumber \\ 
\!&\!=\!&\!\int_0^1\!\left( 
4 \beta \lambda u v - 9 \lambda^2 u^2 v - 4 \alpha \lambda u_{2x} v
+ 4 \alpha \lambda u_x v_x \right) d\lambda  
\nonumber \\ 
\!&\!=\!&\!
2 \beta u v - 3 u^2 v - 2 \alpha u_{2x} v + 2 \alpha u_x v_x, 
\end{eqnarray}
which is $J^{(4)}$ in (\ref{boussinesqconslaws}).
Set $\beta = 1$ at the end of the computations.  
\subsection{Conservation Laws for the Landau-Lifshitz Equation}
\label{applll} 
Without showing full details, we will compute the six density-flux pairs 
(\ref{heisenbergconslaws}) for (\ref{heisenberg}).
The weights for (\ref{heisenberg}) are $W(u) \!=\! W(v) \!=\! W(w) \!=\! a, \; 
W(\alpha) \!=\! W(\beta) \!=\! W(\gamma) \!=\! 2, \; 
W(\frac{\partial}{\partial t}) \!=\! a + 2,$ 
where $a$ is an arbitrary non-negative integer or rational number. 
This example involves three constants $(\alpha, \beta,$ and $\gamma)$ with 
weights which makes the computations quite cumbersome unless we take 
advantage of the multi-uniformity and the cyclic nature of (\ref{heisenberg}).
\vskip 2pt
\noindent
{\bf Step 1}: {\bf Construct the form of the density}
\vskip 2pt
\noindent
Select a small value for $a$, for example $a=\frac{1}{4},$ and compute a 
density of, say, rank $\frac{1}{4}.$
Start from ${\cal {V}} = \{ u, v, w, \alpha, \beta, \gamma \},$ 
i.e.\ the list of dependent variables and the parameters with weight.
Construct the set ${\cal {M}}$ of all non-constant monomials of the 
(selected) rank $\frac{1}{4}$ or less (without derivatives). 
Thus, ${\cal {M}} \!=\! \{ u,v,w \}.$
Next, for each of the monomials in ${\cal {M}},$ 
introduce the appropriate number of $x$-derivatives so that each term 
has rank $\frac{1}{4}.$
No $x$-derivatives are needed and the removal of divergences or 
divergence-equivalent terms is irrelevant. 
Linearly combine the terms in 
${\cal {R}} \!=\! {\cal {S}} \!=\! \{ u,v,w \}$ to get a candidate density:
\begin{equation}
\label{candidaterho1heisenberg}
\rho = c_1 u + c_2 v + c_3 w.
\end{equation}
It suffices to continue with $\rho = c_1 u.$ 
The remaining terms follow by cyclic permutation.
\vskip 3pt
\noindent
{\bf Step 2}: {\bf Compute the undetermined constants $c_i$}
\vskip 2pt
\noindent
Compute $E = -{\rm D}_t \rho = -{\rm D}_t (c_1 u) = - c_1 u_t.$ 
Use (\ref{heisenberg}) to eliminate $u_t.$
Hence,
\begin{equation}
\label{Erho1heisenberg}
E \!=\! -c_1 \Big( v w_{2x} - w v_{2x} + (\gamma - \beta) v w \Big),
\end{equation}
which is the opposite of the right hand side of the first equation in 
(\ref{heisenberg}).
Since $E \!=\! -{\rm D}_t \, \rho \!=\! {\rm D}_x J,$ the expression $E$ 
must be exact.
Obviously ${\cal {L}}^{(0)}_{u(x)}(E) \equiv 0.$
Compute
\begin{eqnarray}
\label{EulerforE1heisenberg}
{\cal {L}}^{(0)}_{v(x)} (E) \!&\!=\!&\! 
\frac{\partial f}{\partial v} (E) 
- {\rm D}_x \frac{\partial f}{\partial v_x} (E)
+ {\rm D}_{x}^2 \frac{\partial f}{\partial v_{2x}} (E) 
\!=\! - c_1 (\gamma - \beta) w, 
\nonumber \\
{\cal {L}}^{(0)}_{w(x)} (E)  \!&\!=\!&\! 
\frac{\partial f}{\partial w} (E) 
- {\rm D}_x \frac{\partial f}{\partial w_x} (E)
+ {\rm D}_{x}^2 \frac{\partial f}{\partial w_{2x}} (E) 
\!=\! - c_1 (\gamma - \beta) v.
\end{eqnarray}
Set the latter expressions identically equal to zero. 
Solve $c_1 (\beta - \gamma) = 0$ for $c_1 \ne 0.$
Set $c_1 = 1$ to get $\rho = u,$ subject to the condition $\beta = \gamma,$ 
which confirms the result in (\ref{heisenbergconslaws}).
\vskip 3pt
\noindent
{\bf Step 3}: {\bf Compute the flux $J$}
\vskip 2pt
\noindent
Substitute $c_1 = 1$ and $\beta = \gamma$ into (\ref{Erho1heisenberg}) 
to get $E = w v_{2x} - v w_{2x}.$
Apply the homotopy operator (\ref{homotopyvectorux}) with 
(\ref{integrandhomotopyvectorux}) to $E.$
In this example ${\bf u}(x) = (u(x), v(x), w(x)).$
Obviously, $I_{u}(E) \!=\!0,$ since there is no explicit occurrence of $u.$
Compute 
\begin{equation} 
\label{IuforEheisenberg}
I_{v}(E) \!=\!
                     v {\cal {L}}^{(1)}_{v(x)} (E) 
+ {\rm D}_x \left(   v {\cal {L}}^{(2)}_{v(x)} (E) \right) 
\!=\! v \frac{\partial E}{\partial v_x}
  - 2 v {\rm D}_x   \left( \frac{\partial E}{\partial v_{2x}} \right)
  + {\rm D}_x \left( v \frac{\partial E}{\partial v_{2x}} \right) 
\!=\! v_x w - v w_x, 
\end{equation}
and, analogously, $I_{w}(E) \!=\! v_x w - v w_x.$
Finally, compute 
\begin{eqnarray}
\label{j1heisenberg}
J \!&\!=\!&\! {\cal {H}}_{u(x)} (E) 
= \int_0^1\! (I_{u}(E) + I_{v}(E) + I_{w}(E))[\lambda {\bf u}] 
\,\frac{d\lambda}{\lambda} 
\nonumber \\ 
\!&\!=\!&\!\int_0^1\! 2 \lambda \left( v_x w - v w_x \right) \, d\lambda 
= v_x w - v w_x, 
\end{eqnarray}
which is $J^{(1)}$ in (\ref{heisenbergconslaws}).
\vskip 5pt
\noindent
To compute density $\rho^{(4)}$ which is quadratic in $u,v,$ and $w,$ 
we would start from rank $\frac{1}{2}$ and repeat the three steps. 
Step 1 would lead to 
$ \rho = c_1 u^2 + c_2 u v + c_3 v^2 + c_4 u w + c_5 v w + c_6 w^2. $ 
Steps 2 and 3 would result in $c_1 = c_3 = c_6 = 1.$
So, $\rho = u^2 + v^2 + w^2$ and $J = 0,$ which agrees with 
$\rho^{(4)}$ and $J^{(4)} = 0 $ in (\ref{heisenbergconslaws}).
\vskip 4pt
\noindent
Starting with rank $1,$ Step 1 would generate a polynomial density with
the 15 terms that are homogeneous of degree 4.
Steps 2 and 3 would result in $\rho^{(5)}$ and $J^{(5)} = 0$ 
in (\ref{heisenbergconslaws}).
\vskip 5pt
\noindent
The computation of $\rho^{(6)}$ and $J^{(6)}$ is cumbersome and long. 
We only indicate the strategy and give partial results.
Notice that $\rho^{(6)}$ would have rank $\frac{5}{2}$ if $a=\frac{1}{4}$ 
and the candidate density would have, amongst others, all homogeneous terms
of degree nine, which is undesirable. 
It turns out that $a=1$ is a better choice to compute $\rho^{(6)},$ which 
then has rank 4.

Start from ${\cal {V}}$ as above.
Construct the set 
${\cal {M}} \!=\! \{ u^4, u^3, u^2, u, v^4, v^3, \cdots, $ $
u^3 v, u^3 w, \cdots, $ $ u^2 v^2, u^2 w^2, \cdots, u^2 v w, v^2 u w, \cdots, 
$ $ \alpha u^2, \beta u^2, $ $ \cdots, \alpha u v, \cdots, \gamma v w, 
\cdots, u v w, u v, \cdots, \gamma w \}$
of all non-constant monomials of rank $4$ or less (without derivatives). 
Next, for each of the 61 monomials in ${\cal {M}},$ 
introduce the needed $x$-derivatives to make each term rank 4.
Use, for example,
\begin{equation}
\label{buildingblocksheisenberg}
\frac{{\partial}^2 u^2}{\partial{x^2}} 
\!=\! 2 u_x^2 + 2 u u_{2x}, 
\;\;
\frac{{\partial} u^3}{\partial{x}} \!=\! 3 u^2 u_x, 
\;\;
\frac{{\partial}^4 u}{\partial{x^4}} = u_{4x}, 
\;\;
\frac{{\partial} u^2 v}{\partial{x}} 
\!=\! 2 u u_x v + u^2 v_x, 
\;\; 
\frac{{\partial} \alpha u}{\partial{x}} 
\!=\! \alpha u_x.
\end{equation}
Ignore the highest-order terms (typically the last terms).
Augment ${\cal {M}}$ with the remaining terms (deleting numerical factors). 
Next, remove all divergences and divergent-equivalent terms to get 
${\cal {S}}$ with 47 terms (not shown).
Linearly combine the terms in ${\cal {S}}$ to get the candidate density:
\begin{eqnarray}
\label{candidaterho6heisenberg}
\rho \!&\!=\!&\! 
c_1 \alpha u^2 + c_2 \beta u^2 + c_3 \gamma u^2 + c_4 u^4 + c_5 \alpha u v 
+ c_6 \beta u v + c_7 \gamma u v + c_8 u^3 v + \cdots + c_{12} u^2 v^2 
\nonumber \\
&& + c_{13} u v^3 + \cdots + c_{22} u^2 v w + c_{23} u v^2 w + \cdots 
+ c_{25} \alpha w^3 + c_{26} \beta w^3 + c_{27} \gamma w^3 + \cdots 
+ c_{29} u v w^2 
\nonumber \\
&& + c_{30} v^2 w^2 + \cdots + c_{33} w^4
+ c_{34} u u_x v + c_{35} u_x v^2 + c_{36} u u_x w 
+ c_{37} u_x v w + c_{38} u_x w^2 + c_{39} u_x^2 
\nonumber \\
&& + c_{40} u v_x w + c_{41} v v_x w + c_{42} v_x w^2 
+ c_{43} u_x v_x + c_{44} v_x^2 + c_{45} u_x w_x  
+ c_{46} v_x w_x + c_{47} w_x^2,
\end{eqnarray}
where only the most indicative terms are explicitly shown.

Compute ${\rm D}_t \rho,$ use (\ref{heisenberg}) to eliminate 
$u_t, v_t, w_t, u_{tx}, v_{tx},$ and $w_{tx},$ and require that 
${\cal {L}}^{(0)}_{u(x)} (E) \!=\! 
{\cal {L}}^{(0)}_{v(x)} (E) \!=\! 
{\cal {L}}^{(0)}_{w(x)} (E) \equiv 0$ 
which leads to a linear system of 121 equations (not shown) for $c_{1}$ 
through $c_{47}.$
Substitute the solution (not shown) into (\ref{candidaterho6heisenberg}) 
to obtain
\begin{eqnarray}
\label{rho6heisenberg}
\rho \!&\!=\!&\! 
(\alpha c_{25} + \beta c_{26} + \gamma c_{27})(u^2 + v^2 + w^2) 
+ \frac{1}{2} c_{30} (u^2 + v^2 + w^2)^2 
\nonumber \\
&& + c_{47} \left( 
u_x^2 + v_x^2 + w_x^2 + (\gamma - \alpha) u^2 + (\gamma - \beta) v^2 \right), 
\end{eqnarray}
where $c_{25}, c_{26}, c_{27}, c_{30},$ and $c_{47}$ are arbitrary. 
Split the density into independent pieces:
\begin{eqnarray}
\label{heisenbergrho6conslaws}
\rho \!&\!=\!&\! u^2 + v^2 + w^2, \quad \rho = (u^2 + v^2 + w^2)^2, 
\nonumber \\
\rho \!&\!=\!&\! u_x^2 + v_x^2 + w_x^2 + 
(\gamma - \alpha) u^2 + (\gamma - \beta) v^2, 
\end{eqnarray}
which are $\rho^{(4)},\rho^{(5)},$ and $\rho^{(6)}$ in 
(\ref{heisenbergconslaws}).
Use the homotopy operator to compute the flux corresponding to 
(\ref{rho6heisenberg}): 
\begin{eqnarray}
\label{j6heisenberg}
J \!&\!=\!&\!
2 c_{47} \Big( 
(v w_x - w v_x) u_{2x} + (w u_x - u w_x) v_{2x} + (u v_x - v u_x) w_{2x}
\Big.
\nonumber \\
&& \! \Big. + (\beta - \gamma) v w u_x + (\gamma - \alpha) u w v_x 
     + (\alpha - \beta) u v w_x \Big).
\end{eqnarray}
$J^{(4)} = J^{(5)} = 0$ since the terms in $c_{25}, c_{26}, c_{27},$ 
and $c_{30}$ all dropped out.
Set $c_{47} = 1$ to get $J^{(6)}$ in (\ref{heisenbergconslaws}).
\subsection{Conservation Laws for the Shallow Water Wave Equations}
\label{applsww} 
The SWW equations (\ref{sww}) admit weights (\ref{swwweightequations}) 
in which $W(h)$ and $W(\Omega)$ are free. 
The fact that (\ref{sww}) is multi-uniform is advantageous. 
Indeed, we can construct a candidate $\rho$ which is uniform in rank for 
one set of weights and, subsequently, use other choices of weights to 
split $\rho$ into smaller pieces. 
This ``divide and conquer" strategy drastically reduces the complexity of 
the computations as was shown in \cite{WHetalbook05}.

The first few densities and fluxes were given in (\ref{swwconslaws}). 
We compute ${\rho}^{(5)}$ and $J^{(5)}.$ 
Note that ${\rho}^{(5)}$ has rank 3 if we select $W(h) \!=\! a \!=\! 1$ and 
$W(\Omega) \!=\! b \!=\! 2.$ 
However, ${\rho}^{(5)}$ has rank 4 if we take $W(h) \!=\! a \!=\! 0$ and 
$W(\Omega) \!=\! b \!=\! 2.$ 
If we set $W(h) \!=\!0,$ and $W(\Omega) \!=\!3$ then ${\rho}^{(5)}$ has rank 7.
So, the trick is to construct a candidate density which is scaling 
homogeneous for a {\it particular} (fixed) choice of $a$ and $b$ in 
(\ref{swwscaleclass}) and split the density based on other choices of $a$ 
and $b.$
\vskip 3pt
\noindent
{\bf Step 1}: {\bf Construct the form of the density}
\vskip 2pt
\noindent
Start from ${\cal {V}} = \{ u, v, \theta, h, \Omega \},$ 
i.e.\ the list of variables and parameters with weights.
Use (\ref{swwscaleclass}) with $a=1, b=2,$ to get ${\cal {M}} \!=\! 
\{ \Omega u, \Omega v,\ldots, u^3, v^3, \ldots, u^2 v, u v^2,\ldots,
u^2, v^2, \ldots, u, v, \theta, h \}$
which has 38 monomials of (chosen) rank 3 or less (without derivatives). 

All terms of rank 3 in ${\cal {M}}$ remain untouched. 
To adjust the rank, differentiate each monomial of rank 2 in ${\cal {M}} $
with respect to $x$ ignoring the highest-order term. 
For example, in $\frac{d u^2}{dx} = 2 u u_x,$ the term can be ignored
since it is a total derivative.
The terms $u_x v$ and $-u v_x$ are divergence-equivalent since 
$\frac{d (u v)}{dx} = u_x v + u v_x.$ 
Keep $u_x v.$
Likewise, differentiate each monomial of rank 2 in ${\cal {M}} $ with 
respect to $y$ and ignore the highest-order term. 

Produce the remaining terms for rank 3 by differentiating the monomials of 
rank 1 in ${\cal {M}} $ with respect to $x$ twice, or $y$ twice, or once 
with respect to $x$ and $y.$ 
Again ignore the highest-order terms.
Augment the set ${\cal {M}} $ with the derivative terms of rank 3 to get
${\cal {R}} = \{ \Omega u, \Omega v, \cdots, u v^2, u_x v, u_x \theta, u_x h, 
\cdots, u_y v, u_y \theta, \cdots, \theta_y h \}$ 
which has 36 terms.

Use the ``divide and conquer'' strategy \cite{WHetalbook05} to select 
from ${\cal {R}}$ the terms which are of ranks 4 and 7 using the choices
$a\!=\!0,b\!=\!2$ and $a\!=\!0, b\!=\!3,$ respectively.
This gives ${\cal {S}} \!=\! 
\{ \Omega \theta, u_x \theta, u_y \theta, v_x \theta, v_y \theta \},$
which happens to be free of divergences and divergence-equivalent terms. 
So, no further reduction is needed. 
Linearly combine the terms in ${\cal {S}}$ to get 
\begin{equation}
\label{swwcandidaterho}
\rho = c_1 \Omega \theta + c_2 u_x \theta  + c_3 u_y \theta + 
c_4 v_x \theta  + c_5 v_y \theta .
\end{equation}
\vskip 2pt
\noindent
{\bf Step 2}: {\bf Compute the undetermined constants $c_i$}
\vskip 2pt
\noindent
Compute $E \!=\! -{\rm D}_t \rho$ and use (\ref{sww}) to remove all time 
derivatives:
\begin{eqnarray}
\label{Erho7sww}
E \!&\!=\!&\! 
- \Big( \frac{\partial \rho}{\partial u_{x}} u_{tx}
  + \frac{\partial \rho}{\partial u_{y}} u_{ty}
  + \frac{\partial \rho}{\partial v_{x}} v_{tx}
  + \frac{\partial \rho}{\partial v_{y}} v_{ty} 
  + \frac{\partial \rho}{\partial \theta } \theta_t \Big)
\nonumber \\
&\!\!=\!\!&\! 
    c_2 \theta (u u_x + v u_y - 2 \Omega v + 
    {\textstyle \frac{1}{2}} h \theta_x + \theta h_x)_x
+\, c_3 \theta (u u_x + v u_y - 2 \Omega v + 
    {\textstyle \frac{1}{2}} h \theta_x + \theta h_x)_y
\nonumber \\
&\!\!\!\!&\! 
+\, c_4 \theta (u v_x + v v_y + 2 \Omega u + 
    {\textstyle \frac{1}{2}} h \theta_y + \theta h_y)_x
+\, c_5 \theta (u v_x + v v_y + 2 \Omega u + 
    {\textstyle \frac{1}{2}} h \theta_y + \theta h_y)_y 
\nonumber \\
&\!\!\!\!&\! 
+\, 
(c_1 \Omega + c_2 u_x + c_3 u_y + c_4 v_x + c_5 v_y)(u \theta_x + v \theta_y).
\end{eqnarray}
Require that 
${\cal {L}}^{(0,0)}_{u(x,y)} (E) 
\!=\! {\cal {L}}^{(0,0)}_{v(x,y)} (E) 
\!=\! {\cal {L}}^{(0,0)}_{\theta(x,y)} (E) 
\!=\! {\cal {L}}^{(0,0)}_{h(x,y)} (E) \equiv 0,$
where, for example, ${\cal {L}}^{(0,0)}_{u(x,y)}$ is given in 
(\ref{zeroeulerscalaruxy}).
Gather like terms. 
Equate their coefficients to zero to obtain 
\begin{equation}
\label{systemsww}
c_1 + 2 c_3 = 0, \quad c_2 = c_5 = 0, \quad
c_1 - 2 c_4 = 0, \quad c_3 + c_4 = 0 .
\end{equation}
Set $c_1 = 2.$ 
Substitute the solution
\begin{equation}
\label{solrho7sww}
c_1 = 2, \; c_2 = 0, \; c_3 = - 1, \; c_4 = 1, \; c_5 = 0.
\end{equation}
into $\rho$ to obtain
\begin{equation}
\label{rho7sww}
\rho = 2 \Omega \theta - u_y \theta + v_x \theta,
\end{equation}
which is $\rho^{(5)}$ in (\ref{swwconslaws}).
\vskip 3pt
\noindent
{\bf Step 3}: {\bf Compute the flux ${\bf J}$}
\vskip 2pt
\noindent
Compute the flux corresponding to (\ref{rho7sww}).
To do so, substitute (\ref{solrho7sww}) into (\ref{Erho7sww}) to obtain 
\begin{eqnarray}
\label{Erho7swwupdate}
E \!&\!=\!&\!
\theta ( u_x v_x - u_y v_y + u v_{2x} - u_{2y} v
- u_x u_y + v_x v_y - u u_{xy}  + v v_{xy} 
\nonumber \\
\!&\!\!&\!
+ 2 \Omega u_x + 2 \Omega v_y 
+ {\textstyle \frac{1}{2}} \theta_x h_y
- {\textstyle \frac{1}{2}} \theta_y h_x ) 
+ 2 \Omega u \theta_x + 2 \Omega v \theta_y - u u_y \theta_x 
\nonumber \\
\!&\!\!&\! - u_y v \theta_y + u v_x \theta_x + v v_x \theta_y .
\end{eqnarray}
Apply the 2D homotopy formulas in 
(\ref{homotopyvectoruxycompx})-(\ref{integrandhomotopyvectoruxycompy})
to $E = {\rm Div} \, {\bf J} = {\rm D}_x J_1 + {\rm D}_y J_2.$ 
So, compute 
\begin{eqnarray}
\label{Iuforsww}
\nonumber
I_u^{(x)}(E) 
\!&\!=\!\!&\! u {\cal {L}}^{(1,0)}_{u(x,y)}(E)
+ {\rm D}_x \left( u {\cal {L}}^{(2,0)}_{u(x,y)}(E) \right)
+ \frac{1}{2}{\rm D}_y \left( u {\cal {L}}^{(1,1)}_{u(x,y)}(E) \right)
\\ \nonumber
\!&\!=\!&\!
u \left( \frac{\partial E}{\partial u_x}
\!-\!2{\rm D}_x \left( \frac{\partial E}{\partial u_{2x}} \right)
\!-\!{\rm D}_y \left( \frac{\partial E}{\partial u_{xy}} \right) \right)
\!+\!{\rm D}_x \left(\! u \frac{\partial E}{\partial u_{2x}} \!\right)
\!+\!\frac{1}{2}{\rm D}_y 
\left(\! u \frac{\partial E}{\partial u_{xy}} \!\right)
\\
\!&\!=\!&\!
u v_x \theta + 2 \Omega u \theta + \frac{1}{2} u^2 \theta_y - u u_y \theta.
\end{eqnarray}
Similarly, compute
\begin{eqnarray}
\label{Ivforsww}
I_v^{(x)}(E) &=& v v_y \theta + \frac{1}{2} v^2 \theta_y + u v_x \theta,
\\
I_{\theta}^{(x)}(E) &=&
\frac{1}{2} \theta^2 h_y + 2 \Omega u \theta - u u_y \theta + u v_x \theta, 
\\
I_{h}^{(x)}(E) &=&
- \frac{1}{2} \theta \theta_y h.
\end{eqnarray}
Next, compute 
\begin{eqnarray}
\label{dellarhomotop2Dxpart}
J_1 ({\bf u}) 
\!&\!=\!&\! {\cal {H}}^{(x)}_{{\bf u}(x,y)} (E) 
\nonumber \\
\!&\!=\!&\!
\int_{0}^{1}\! \left( I_u^{(x)}(E) + I_v^{(x)}(E) + I_{\theta}^{(x)}(E) 
+ I_h^{(x)}(E) \right)[\lambda {\bf u}] \,\frac{d \lambda}{\lambda}
\nonumber \\
\!&\!=\!&\!
\int_0^1\! \left( 4 \lambda \Omega u \theta + 
\lambda^2 \left( 
3 u v_x \theta + \frac{1}{2} u^2 \theta_y - 2 u u_y \theta
+ v v_y \theta + \frac{1}{2} v^2 \theta_y 
\right. \right.
\nonumber \\
\!&\!&
\left. \left.
\;\;\;\;\;\, + \frac{1}{2} \theta^2 h_y - \frac{1}{2} \theta \theta_y h
\right) \right) d\lambda  
\nonumber \\ 
\!&\!=\!&\!
2 \Omega u \theta - \frac{2}{3} u u_y \theta +  u v_x \theta 
+ \frac{1}{3} v v_y \theta  + \frac{1}{6} u^2 \theta_y 
+ \frac{1}{6} v^2 \theta_y - \frac{1}{6} h \theta \theta_y 
+ \frac{1}{6} h_y \theta^2.
\end{eqnarray}
Likewise, compute $I_u^{(y)}(E), I_v^{(y)}(E), I_{\theta}^{(y)}(E),$
and $I_h^{(y)}(E).$
Finally, compute
\begin{eqnarray}
J_2 ({\bf u}) 
\!&\!=\!&\! {\cal {H}}^{(y)}_{{\bf u}(x,y)} (E) 
\nonumber \\ 
\!&\!=\!&\!
\int_{0}^{1}\! \left( I_u^{(y)}(E) + I_v^{(y)}(E) + I_{\theta}^{(y)}(E) 
+ I_h^{(y)}(E) \right)[\lambda {\bf u}] \,\frac{d \lambda}{\lambda}
\nonumber \\
\! &\!=\!& \!
2 \Omega v \theta + \frac{2}{3} v v_x \theta - v u_y \theta 
- \frac{1}{3} u u_x \theta  - \frac{1}{6} u^2 \theta_x 
- \frac{1}{6} v^2 \theta_x 
+ \frac{1}{6} h \theta \theta_x - \frac{1}{6} h_x \theta^2.
\end{eqnarray}
Hence, 
\begin{equation}
\label{J7sww} 
{\bf J} = \!\left(\! \begin{array}{c}  J_1 \\ J_2 \end{array} \!\right)\!
= \frac{1}{6} \left(
 \begin{array}{c}
 12 \Omega u \theta - 4 u u_y \theta + 6 u v_x \theta + 2 v v_y \theta 
 + u^2 \theta_y + v^2 \theta_y - h \theta \theta_y + h_y \theta^2 
 \\
 12 \Omega v \theta + 4 v v_x \theta - 6 v u_y \theta - 2 u u_x \theta 
 - u^2 \theta_x - v^2 \theta_x + h \theta \theta_x - h_x \theta^2 
 \end{array}
 \right), 
\end{equation}
which matches ${\bf J}^{(5)}$ in (\ref{swwconslaws}).
\section{Conclusions}
\label{conclusions}
%
%
The variational derivative (zeroth Euler operator) provides a 
straightforward way to test exactness which is key in the computation of 
densities.
The continuous homotopy operator is a powerful, algorithmic tool to compute 
fluxes explicitly. 
Indeed, the homotopy operator allows one to invert the total divergence 
operator by computing higher variational derivatives followed by a 
one-dimensional integration with respect to a single auxiliary parameter.
The homotopy operator is a universal tool that can be applied to problems 
in which integration by parts (of arbitrary functions) in multi-variables 
is crucial.

To reach a wider audience, we intentionally did not use differential forms 
and the abstract framework of variational bi-complexes and homological 
algebra. 
Instead, we extracted the Euler and homotopy operators from their abstract 
setting and presented them in the language of standard calculus, thereby
making them widely applicable to computational problems in the sciences.
Our calculus-based formulas can be readily implemented in CAS.

Based on the concept of scaling invariance and using tools of the calculus 
of variations, we presented a three-step algorithm to symbolically compute 
polynomial conserved densities and fluxes of nonlinear polynomial systems 
of PDEs in multi-spatial dimensions.
The steps are straightforward: build candidate densities as linear 
combinations (with undetermined constant coefficients) of terms 
that are homogeneous with respect to the scaling symmetry of the PDE. 
Subsequently, use the variational derivative to compute the coefficients,
and, finally, use the homotopy operator to compute the flux. 

With our method one can search for conservation laws in chemistry, physics, 
and engineering. 
Symbolic packages are well suited to assist in the search which covers many 
fields of application including fluid mechanics, plasma physics, 
electro-dynamics, gas dynamics, elasticity, nonlinear optics and
acoustics, electrical networks, chemical reactions, etc.. 
\section*{Acknowledgements and Dedication}
\label{acknowledgements}
This material is based upon work supported by the National Science 
Foundation (NSF) under Grants Nos.\ DMS-9732069, DMS-9912293, and 
CCR-9901929. 
Any opinions, findings, and conclusions or recommendations expressed in this
material are those of the author and do not necessarily reflect the views 
of NSF. 

The author is grateful to Bernard Deconinck, Michael Colagrosso, Mark Hickman 
and Jan Sanders for valuable discussions. 
Undergraduate students Lindsay Auble, Robert ``Scott" Danford, Ingo Kabirschke,
Forrest Lundstrom, Frances Martin, Kara Namanny, Adam Ringler, and Maxine 
von Eye are thanked for designing {\it Mathematica} code for the project.  

The research was supported in part by an Undergraduate Research Fund Award 
from the Center for Engineering Education awarded to Ryan Sayers.
On June 16, 2003, while rock climbing in Wyoming, Ryan was killed by a 
lightning strike. 
He was 20 years old. 
The author expresses his gratitude for the insight Ryan brought 
to the project.
His creativity and passion for mathematics were inspiring.
This paper is dedicated to him.


\begin{thebibliography}{99}
%
%
\bibitem{MAandPCbook91}
Ablowitz, M.\ J.; Clarkson, P.\ A.\
{\rm Solitons, Nonlinear Evolution Equations and Inverse Scattering};
Cambridge University Press: Cambridge, 1991.

\bibitem{MAandHSbook81}
Ablowitz, M.\ J.; Segur, H.\
{\rm Solitons and the Inverse Scattering Transform};
SIAM: Philadelphia, 1981.

\bibitem{PAthesis03}
Adams, P.\ J.\
{\rm Symbolic Computation of Conserved Densities and Fluxes for Systems of 
Partial Differential Equations with Transcendental Nonlinearities};
MS Thesis, Dept.\ Math.\ \& Comp.\ Scs.;
Colorado School of Mines: Golden, Colorado, 2003.


\bibitem{SA03} 
Anco, S.\ C.\
J Phys A: Math Gen 2003, 36, 8623-8638.

\bibitem{SAandGB02}
Anco, S.\ C.; Bluman, G.\
Euro J Appl Math 2002, 13, 545-566 \& 567-585.



\bibitem{IAbook04}
Anderson, I.\ M.\
{\rm The Variational Bicomplex}; 
Dept.\ Math., Utah State University: Logan, Utah,
November 2004, 318 pages;
\newline
\verb|http://www.math.usu.edu/~fg_mp/Publications/VB/vb.pdf.|

\bibitem{IAvessiothandbook04}
Anderson, I.\ M.\
{\rm The Vessiot Package}, 
Dept.\ Math., Utah State University: Logan, Utah,  
November 2004, 253 pages;
\newline
\verb|http://www.math.usu.edu/~fg_mp/Pages/SymbolicsPage/VessiotDownloads.html|


\bibitem{MBetal04}
Barnett, M.\ P.; Capitani, J.\ F.; von zur Gathen J.; Gerhard, J.\
Int J Quan Chem 2004, 100, 80-104.

\bibitem{JCetal02}
Cantarella, J.; DeTurck, D.; Gluck, H.\
Am Math Monthly 2002, 109, 409-442.


\bibitem{BDwebsite04}
Deconinck, B.; Nivala, M.\
{\rm A {\rm Maple} Package for the Computation of Conservation Laws};
\newline
\verb|http://www.amath.washington.edu/~bernard/papers.html.|

\bibitem{PDpf03}
Dellar, P.\
Phys Fluids 2003, 15, 292-297.


\bibitem{HEthesis03}
Eklund, H.\
{\rm Symbolic Computation of Conserved Densities and Fluxes for 
Nonlinear Systems of Differential-Difference Equations};
MS Thesis, Dept.\ Math.\ \& Comp.\ Scs.;
Colorado School of Mines: Golden, Colorado, 2003.

\bibitem{LFandLT87}
Faddeev, L.\ D.; Takhtajan, L.\ A.\
{\rm Hamiltonian Methods in the Theory of Solitons};
Springer Verlag: Berlin, 1987.

\bibitem{MGetalcpc02}
Gao, M.; Kato, Y.; Ito, M.\
Comp Phys Comm 2002, 148, 242-255.

\bibitem{MGetal04} 
Gao, M.; Kato, Y.; Ito, M.\
Comp Phys Comm 2004, 160, 69-89.

\bibitem{UGandWHjsc97}
G\"{o}kta\c{s}, \"{U}.; Hereman, W.\
J Symb Comp 1997, 24, 591-621.

\bibitem{UGandWHpd98}
G\"{o}kta\c{s}, \"{U}.; Hereman, W.\
Phys D 1998, 132, 425-436.


\bibitem{WHwebsite04}
Hereman, W.\ 
{\rm Mathematica} Packages for the Symbolic Computation of Conservation Laws 
of Partial Differential Equations and Differential-difference Equations;
\newline
\verb|http://www.mines.edu/fs_home/whereman/.| 

\bibitem{WHetalbook05}
Hereman, W.; Colagrosso, M.; Sayers, R.; Ringler, A.; Deconinck, B.;
Nivala, M.; Hickman, M.\ S.\
In Differential Equations with Symbolic Computation;
Wang, D.; Zheng, Z., Eds.; Birkh\"auser Verlag: Basel, 2005. 

\bibitem{WHetalcrm04} 
Hereman, W.; Sanders, J.\ A.; Sayers, J.; Wang, J.\ P.\
In {\rm Group Theory and Numerical Methods},  
CRM Proc.\ Lect.\ Ser.\ {\bf 39},
Winternitz, P.; Gomez-Ullate, D., Eds.;
AMS: Providence, Rhode Island, 2005, pp.\ 267-282.


\bibitem{MHandWHprsa03} 
Hickman, M.; Hereman, W.\
Proc Roy Soc Lond A 2003, 459, 2705-2729.





\bibitem{IKandAVbook98}
Krasil'shchik, I.\ S.; Vinogradov, A.\ M.\
{\rm Symmetries and Conservation Laws for Differential Equations 
of Mathematical Physics}.
AMS: Providence, Rhode Island, 1998.

\bibitem{MKetal70} 
Kruskal, M.\ D.; Miura, R.\ M.; Gardner, C.\ S., Zabusky; N.\ J.\ 
J Math Phys 1970, 11, 952-960.





\bibitem{JMandATbook88}
Marsden, J.\ E.; Tomba, A.\ J.\
{\rm Vector Calculus}, 3rd ed.; 
W.\ H.\ Freeman and Company: New York, 1988.

%



\bibitem{AN83}
Newell, A.\ C.\
J Appl Mech 1983, 50, 1127-1137.

\bibitem{PObook93}
Olver, P.\ J.\
{\rm Applications of Lie Groups to Differential Equations},
2nd ed.;
Springer Verlag: New York, 1993.

\bibitem{JMS1982}
Sanz-Serna, J.\ M.\
J Comput Phys 1982, 47, 199-210.

\bibitem{MSandHF93}
Svendsen, M.; Fogedby, H.\ C.\
J Phys A: Math Gen 1993, 26, 1717-1730.



\bibitem{TW02}
Wolf, T.\
Euro J Appl Math 2002, 13, 129-152.

\bibitem{TWetal03} 
Wolf, T.; Brand, A.; Mohammadzadeh, M.\
J Symb Comp 1999, 27, 221-238.
%
\end{thebibliography}
\end{document}